\documentclass[10pt]{article}

\usepackage{xspace}

\usepackage{epsfig}
\usepackage{graphics}
\usepackage{graphicx}
\usepackage{amsmath}
\usepackage{amsfonts}
\usepackage{amsthm}
\usepackage{amssymb}
\usepackage{url}
\usepackage{subfigure}
\usepackage{multicol}
\usepackage{captcont}
\usepackage{setspace}

\usepackage{fancyhdr}
\newtheorem{finding}{Finding}

\newcommand{\One }{Beijing\xspace}
\newcommand{\Two }{Seoul\xspace}
\newcommand{\Three }{Tokyo\xspace}
\newcommand{\Four }{Hong Kong\xspace}
\newcommand{\Five }{Ha Noi\xspace}
\newcommand{\Six }{Ho Chi Minh\xspace}
\newcommand{\Seven }{Jakarta\xspace}
\newcommand{\Eight }{Kolkata\xspace}
\newcommand{\Nine }{Mumbai\xspace}
\newcommand{\Ten }{Delhi\xspace}
\newcommand{\Eleven }{Karachi\xspace}
\newcommand{\Twelve }{Tehran\xspace}
\newcommand{\Thirteen }{Moscow\xspace}
\newcommand{\Fourteen }{Istanbul\xspace}
\newcommand{\Fifteen }{London\xspace}
\newcommand{\Sixteen }{Lagos\xspace}
\newcommand{\Seventeen }{Kinshasa\xspace}
\newcommand{\Eighteen }{New York\xspace}
\newcommand{\Nineteen }{Mexico City\xspace}
\newcommand{\Twenty }{Bogota\xspace}
\newcommand{\Twentyone }{Lima\xspace}
\newcommand{\Twentytwo }{Sao Paulo\xspace}
\newcommand{\Twentythree }{Canberra\xspace}
\newcommand{\Twentyfour }{Wellington\xspace}

\title{\bf The World's Colonisation and Trade Routes Formation as Imitated by Slime Mould}

\author{Andrew Adamatzky \\
University of the West of England, Bristol, United Kingdom\\ 
\vspace{0.5cm}\\ 
This is unedited preprint with low-resolution photographs.\\  
Final and edited version of this paper is published in \\ \bf
 Int. J. Bifurcation Chaos, 22, 1230028 (2012) [26 pages] \\ 
DOI: 10.1142/S0218127412300285
}

\date{}

\begin{document}

\maketitle


\begin{abstract}
The plasmodium of  \emph{Physarum polycephalum} is renowned for spanning sources of nutrients with 
networks of protoplasmic tubes. The networks transport nutrients and metabolites across the plasmodium's 
body. To imitate a hypothetical colonisation of the world and formation of major transportation routes we cut continents from agar plates arranged in Petri dishes or on the surface of a three-dimensional globe, represent positions of selected metropolitan areas with oat flakes and inoculate the plasmodium in one of the metropolitan areas.  The plasmodium propagates towards the sources of  nutrients, spans them with its network of protoplasmic tubes and even crosses bare substrate between the continents. From the laboratory experiments we derive weighted Physarum graphs, analyse their structure, compare them with the basic proximity graphs and generalised graphs derived from the Silk Road and the Asia Highway networks.

\vspace{0.5cm}

\noindent
\emph{Keywords: biological transport networks, unconventional computing, slime mould} 
\end{abstract}

\section{Introduction}

Nature-inspired computing paradigms and experimental laboratory prototypes are 
demonstrated reasonable success in approximation of shortest, and often collision-free, paths between two given points 
in an Euclidean space or a graph. Examples include ant-based optimisation of communication 
networks~\cite{Dorigo_2004}, approximation of a shortest path in experimental reaction-diffusion chemical systems~\cite{adamatzky_2005}, gas-discharge analog systems~\cite{reyes_2002}, spatially extended crystallisation 
systems~\cite{adamatzky_hotice}, fungi mycelia networks~\cite{jarret_2006}, and maze solving by \emph{Physarum polycephalum}~\cite{nakagaki_2001}.  Amongst all explored so far experimental prototypes of path-computing devices slime mould \emph{P. polycephalum} is the most user-friendly and easiest to cultivate and observe biological substrate. 
 In its plasmodium stage, \emph{P. polycephalum} spans scattered sources of nutrients with a network of 
 protoplasmic tubes.  Nakagaki and colleagues demonstrated that the plasmodium optimises its protoplasmic network to cover all sources of nutrients and to assure quick and fault-tolerant distribution of nutrients in the plasmodium's 
 body~\cite{nakagaki_2000, nakagaki_2001, nakagaki_iima_2007, nakagaki_2007}. There are experimental evidences~\cite{adamatzky_ppl_2008} that in the course of the plasmodium's foraging behaviour the protoplasmic network follows the Toussaint~\cite{toussaint_1980,matula_sokal_1984,jaromczyk_toussaint_1992} hierarchy of proximity graphs however neither of the known proximity graphs is exactly matched by the protoplasmic networks~\cite{adamatzky_physarummachines}.

The slime mould's protoplasmic network is responsible for transportation of nutrients and metabolites, 
and intra-cellular communication. Therefore it is a matter of the natural scientific curiosity to compare the protoplasmic 
transport networks with man-made transport network. Two first comparisons were made by Adamatzky and 
Jones~\cite{adamatzky_jones_2009} --- the slime mould imitation of the UK motorways,  and Tero and colleagues \cite{tero_2010} --- the slime mould imitation of the Tokyo rail roads. These pioneering results were followed by a series of laboratory experiments on evaluation of motorway/highway networks in Mexico~\cite{adamatzky_mexican}, 
Spain and Portugal~\cite{adamatzky_alonso_2010},  Brazil~\cite{adamatzky_pedro}, 
Australia~\cite{adamatzky_prokopenko} and Canada~\cite{adamatzky_akl}. These papers demonstrated  
that \emph{P. polycephalum} approximates existing road networks up to some degree, especially with regards to the core spanning structure, but the slime mould also allows for a wide range of structural deviations in the transport networks and solutions are not always optimal or predictable. The shape of any particular country and the spatial configuration of major urban areas, represented by sources of nutrients, might substantially determine the resultant topology of the protoplasmic networks developed.  In certain countries, the road network designs were influenced not only by logic but also by political, economic and somewhat idealogical decisions. Therefore, more experiments 
are required to provide a viable generalisation, to undertake a comparative analyses of slime mould transport networks 
constructed in different countries and to built a theory of \emph{P. polycephaum} based road planning and associated urban development.  In present paper we decided to undertake an experimental laboratory exercise on playing a hypothetical scenario of the colonisation of the World by large-scale amorphous substrate and formation of the world-wide transportation routes crossing countries and linking continents. The resultant protoplasmic networks were compared, at the abstract level of generalised graphs, with ancient road network --- the Silk Road, and future road network --- the Asian Highways.

\section{Materials and methods}
\label{methods}

The plasmodium of \emph{ P. polycephalum} is cultivated in a plastic container, on paper kitchen towels moistened with 
still water, and fed with oat flakes. For experiments we use $220 \times 220$~mm polystyrene square Petri dishes
and 2\% agar gel (Select agar, by Sigma Aldrich) as a substrate and also the three-dimensional globe 15~cm diameter 
(Stellanova, Germany). When experiments are conducted in the Petri dishes the hot agar gel is poured in the dishes 
to fill the dishes by 2--3~mm. When experimenting with the globe we poured hot gel onto the globe while rotating the globe, so it is covered by agar uniformly. The globe was covered by the agar gel layer by layer. When one layer cooled down and settled, next layer of hot get applied.  When agar cooled down, the shapes corresponding to oceans and seas are cut out and removed, and only the dry land is represented by the agar substrate.

\begin{figure}[!tbp]
\centering
\subfigure[]{\includegraphics[width=0.8\textwidth]{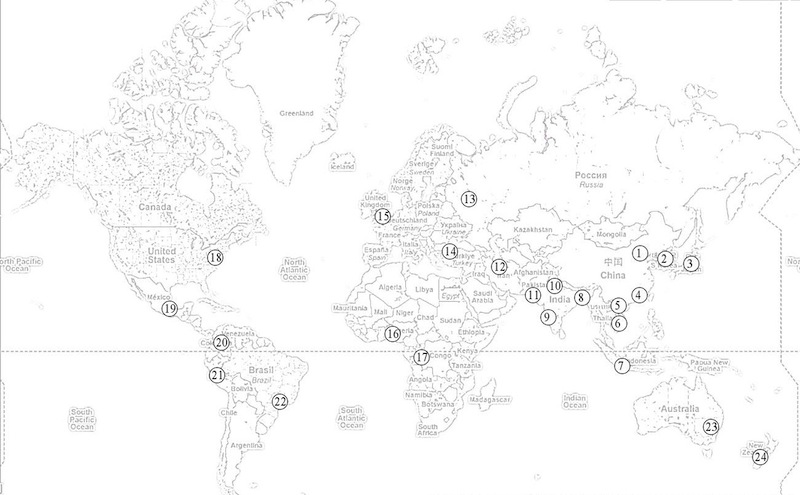}}
\subfigure[]{\includegraphics[width=0.8\textwidth]{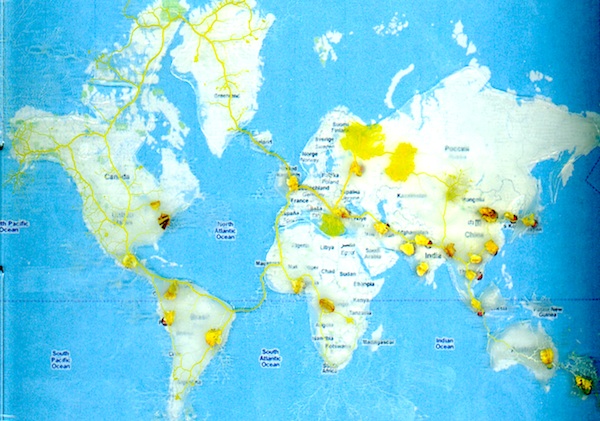}}
\caption{Experimental setup. (a)~Configuration of points, representing selected metropolitan or urban areas $\mathbf{U}$.
(b)~Slime mould \emph{P. polycephalum} occupies oat flakes representing urban areas $\mathbf{U}$.}
\label{setup}
\end{figure}

\begin{figure}[!tbp]
\centering
\includegraphics[width=0.95\textwidth]{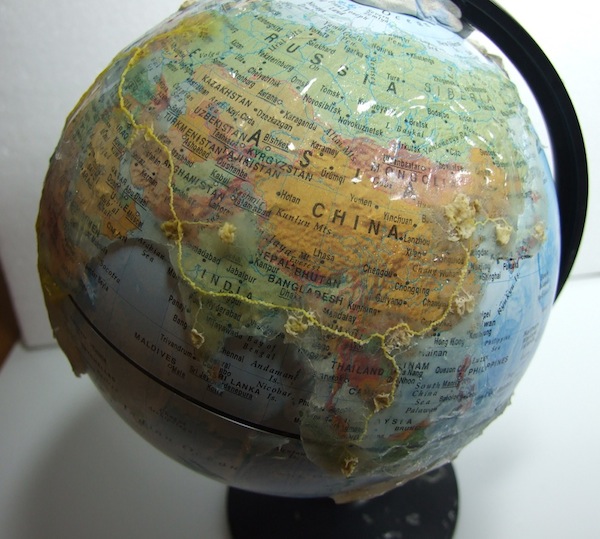}
\caption{Experimental setup. Globe covered with agar gel is colonised by slime mould \emph{P. polycephalum}. Oat flakes represent areas of $\mathbf{U}$.}
\label{globeExm}
\end{figure}

\begin{figure}[!tbp]
\centering
\subfigure[]{\includegraphics[width=0.47\textwidth]{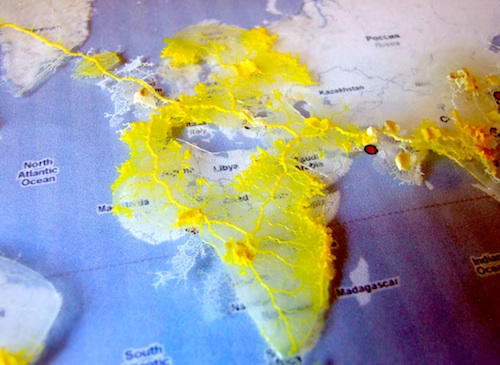}}
\subfigure[]{\includegraphics[width=0.51\textwidth]{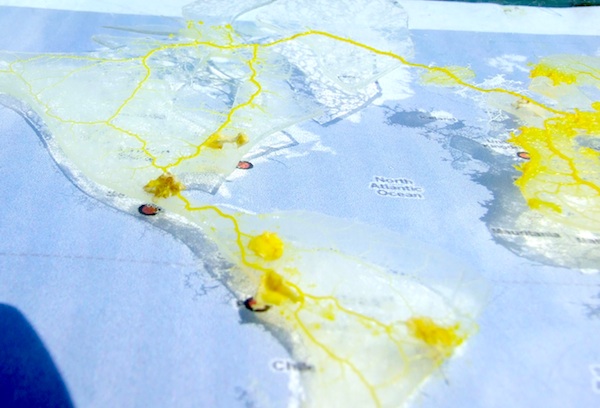}}
\caption{Plasmodium crosses bare space between the continent-shaped agar plates.}
\label{fotos2D}
\end{figure}

We considered 24 metropolitan areas as follows (Fig.~\ref{setup}a): 
\begin{multicols}{5}
\begin{enumerate}
\item Beijing 
\item Seoul 
\item Tokyo 
\item Hong Kong 
\item Ha Noi 
\item Ho Chi Minh 
\item Jakarta 
\item Kolkata 
\item Mumbai 
\item Delhi 
\item Karachi 
\item Tehran 
\item Moscow 
\item Istanbul 
\item London 
\item Lagos 
\item Kinshasa 
\item New York 
\item Mexico City 
\item Bogota 
\item Lima 
\item Sao Paulo
\item Canberra 
\item Wellington 
\end{enumerate}
\end{multicols}
The elements of $\mathbf{U}$ were selected based on their population size, proximity to other urban/metropolitan areas, representativeness of a continent/country. The following area were not chosen for experiments, despite being amongst largest metropolitan areas:  Manila is not included due to its proximity to Jakarata; Shanghai due to its proximity to Beijing; Buenos Aires due to its proximity to  San Paolo; Osaka-Kobe-Kyoto due their proximity to Tokyo; Dhaka due to its proximity to Kalkutta. The capital cities Canberra and Wellington are by no means in the list of largest cities or metropolitan areas, however they were included to give slime mould a change to move to these countries. 
To represent areas of $\mathbf{U}$ we placed oat flakes (each flake weights 9--13~mg and is 5--7~mm in diameter) 
in the positions of agar plate corresponding to the areas (Figs.~\ref{setup}b and \ref{globeExm}). 

At the beginning of each experiment an oat flake colonised by plasmodium (25--30~mg plasmodial weight) is placed in Beijing area.   We have chosen Beijing as a starting point for the slime mould  because Beijing is one of the most populated cities in the world~\cite{Beijing} and also because the Chinese civilisation is amongst earliest  
civilisations~\cite{makeham_2008,makeham_2008}.

We undertook 38 experiments in total: eight experiments with the globe and 30 experiments with the Petri dish.  The Petri dishes  and the globe with plasmodium were kept in darkness (the globe was placed in a glass container to prevent drying of agar gel), at temperature  22-25$^\text{o}$C, except for observation and image recording. Periodically, the dishes were scanned with an Epson  Perfection 4490 scanner and the globe was photographed with FujiFilm FinePix. As  exemplified in Fig.~\ref{fotos2D} absence of a humid growing substrate prevented plasmodium from spreading 'uncontrollably' into parts of substrate corresponding to oceans and seas yet allowed the plasmodium to migrate 
between the continents when necessary. 

To generalise our experimental results we constructed a Physarum graph with weighted-edges.
A Physarum graph  is a tuple ${\mathbf P} = \langle {\mathbf U}, {\mathbf E}, w  \rangle$,
where $\mathbf U$ is a set of  urban areas, $\mathbf E$ is a set edges, and
$w: {\mathbf E} \rightarrow [0,1]$ associates each edge of $\mathbf{E}$ with  a probability (or weights).
For every two regions $a$ and $b$ from $\mathbf U$ there is an edge connecting $a$ and $b$ if a
plasmodium's protoplasmic link is recorded at least in one of $k$ experiments, and the edge $(a,b)$ has a
probability calculated as a ratio of experiments where protoplasmic link $(a,b)$ occurred in the total number
of experiments $k$. We do not take into account the exact configuration of the protoplasmic tubes but merely their existence. Further we will be dealing with threshold Physarum graphs $\mathbf{P}(\theta)  = \langle  {\mathbf U}, T({\mathbf E}), w, \theta \rangle$. The threshold Physarum graph is obtained from Physarum graph by the transformation: $T({\mathbf E})=\{ e \in {\mathbf E}: w(e) \geq \theta \}$. That is all edges with weights less
than $\theta$ are removed.

\section{Results}

\subsection{Scenarios of colonisation}
\label{scenarios}

\begin{figure}[!tbp]
\centering
\subfigure[1 day]{\includegraphics[width=0.49\textwidth]{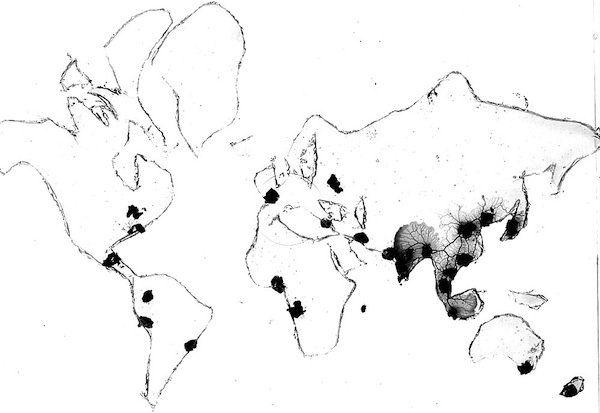}}
\subfigure[2 days]{\includegraphics[width=0.49\textwidth]{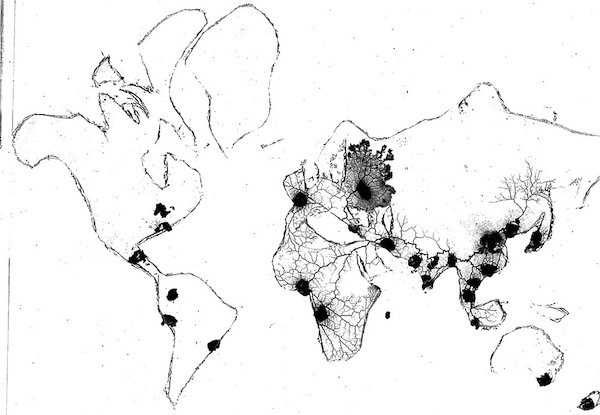}}
\subfigure[3 days]{\includegraphics[width=0.49\textwidth]{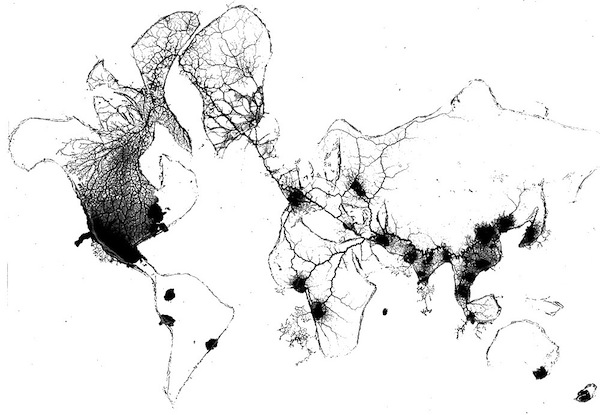}}
\subfigure[4 days]{\includegraphics[width=0.49\textwidth]{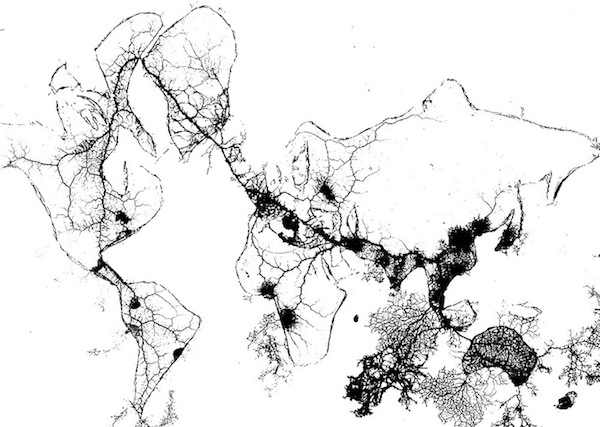}}
\subfigure[5 days]{\includegraphics[width=0.49\textwidth]{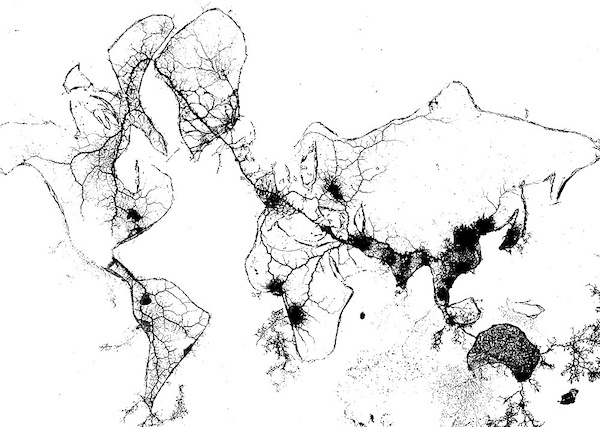}}
\caption{Example of plasmodium development on a configuration continent-shaped agar plates.}
\label{X04}
\end{figure}

Scenarios of the plasmodium development in Petri dishes and on the globe are strikingly similar. 
Here we consider one example of  colonisation of $\mathbf{U}$ in the Petri dish and three 
examples of the colonisation on the globe.  In first day after inoculation in the Petri dish, the 
plasmodium propagates from \One to \Two and \Three in the east; from \One to \Four, \Five and \Six in the south; and from \One to \Eight, \Nine, \Ten and \Eleven in the west (Fig.~\ref{X04}a). On second day the plasmodium explores the space north of \One, links \Eleven and \Twelve, and \Twelve and \Fourteen with protoplasmic tubes. It also growth from \Fourteen  to \Thirteen  and from \Thirteen  to \Fifteen , and from \Twelve  to \Sixteen  and \Seventeen  (Fig.~\ref{X04}b). The plasmodium grows from \Fifteen to Iceland, then to Greenland. It propagates from Greenland to Canada and eventually reaches \Eighteen and \Nineteen on the fifth day of the experiment (Fig.~\ref{X04}c). The final stage of spanning $\mathbf{U}$ takes place on the sixth day of the experiment: the plasmodium propagates from \Nineteen to \Twenty and from \Twenty to \Twentyone and \Twentytwo. At the same time the plasmodium grows from \Six to \Seven and from \Seven to \Twentythree and \Twentyfour (Fig.~\ref{X04}d). As soon as all sources of nutrients, representing the areas of $\mathbf{U}$, are spanned by the plasmodium, the plasmodium remains in its configuration for a couple of days (Fig.~\ref{X04}c). By that time the nutrients become depleted, and the substrate  contaminated with products of metabolism and the plasmodium tries to migrate to other areas and/or form sclerotium.

\begin{figure}[!tbp]
\centering
\subfigure[]{\includegraphics[width=0.8\textwidth]{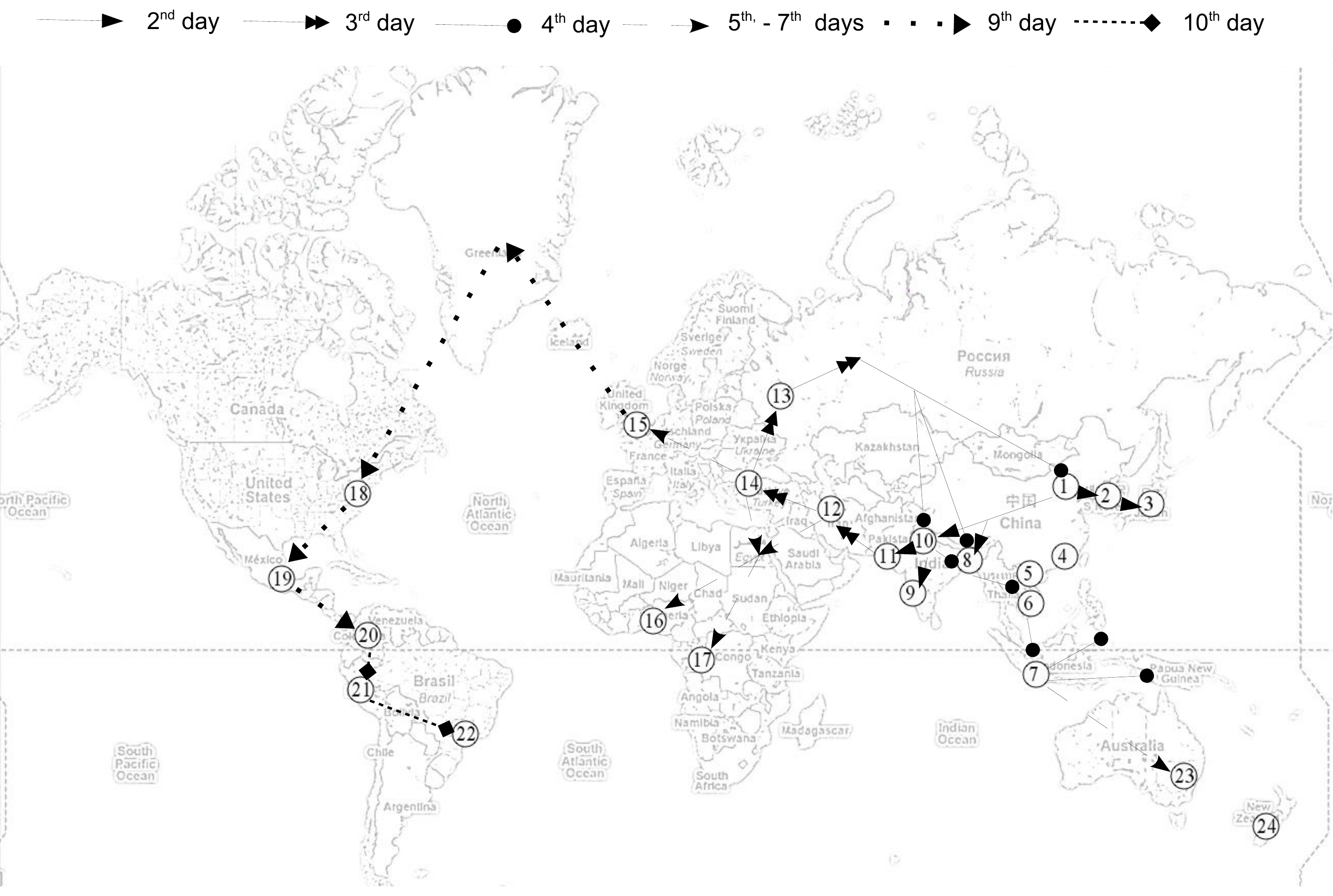}}
\subfigure[]{\includegraphics[width=0.8\textwidth]{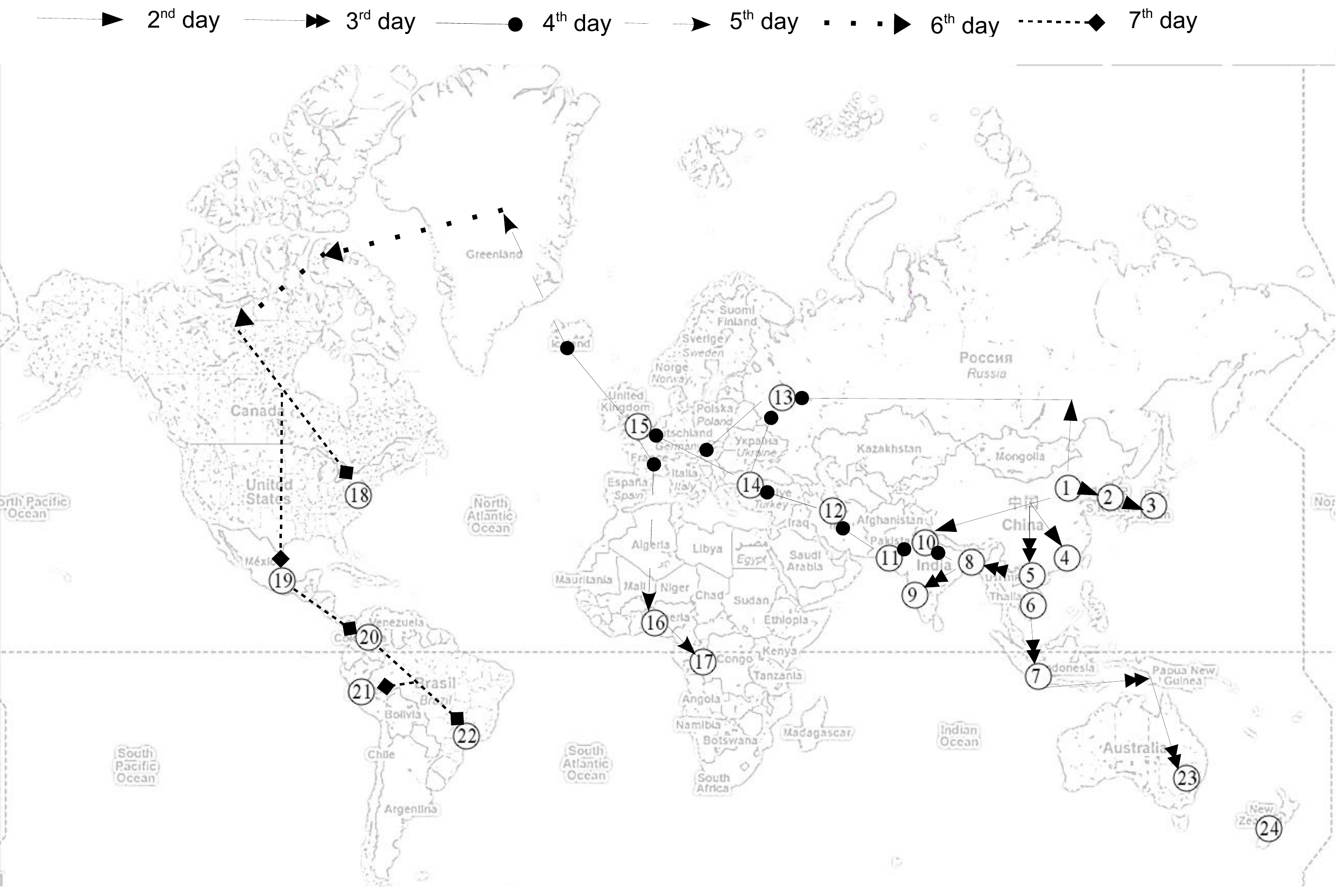}}
\caption{Two scenarios of plasmodial active zones propagation on the globe.}
\label{GlobesTrajectories}
\end{figure}

\begin{figure}[!tbp]
\centering
\subfigure[]{\includegraphics[width=0.49\textwidth]{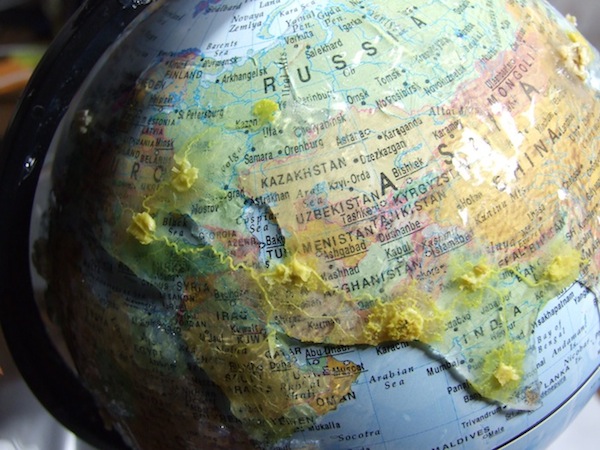}\label{6417}}
\subfigure[]{\includegraphics[width=0.49\textwidth]{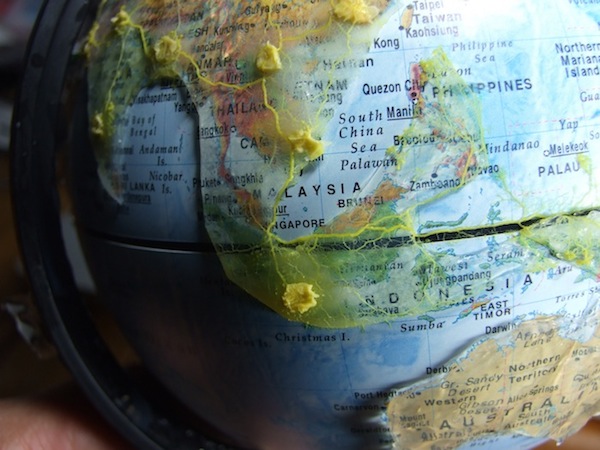}\label{6441}}
\subfigure[]{\includegraphics[width=0.49\textwidth]{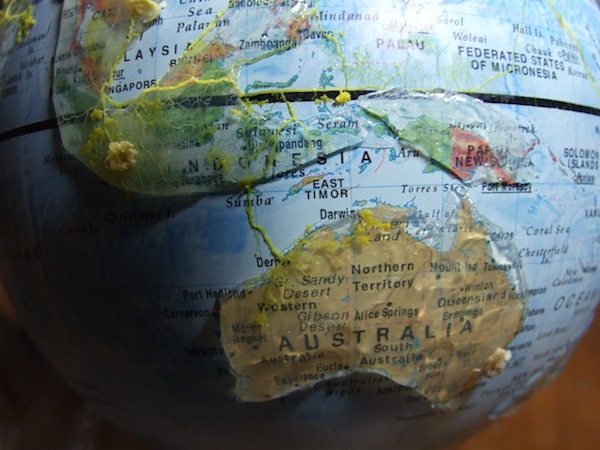}\label{6455}}
\caption{Experimental examples of scenario Fig.~\ref{GlobesTrajectories}a.
(a)~Propagation of slime mould from \Eleven to \Twelve to \Fourteen to \Thirteen on the third day after inoculation.
(b)~On the fourth day of experiment plasmodium links \Five to \Six to \Seven with its protoplasmic tubes and explores space east of \Seven.
(c)~Plasmodium's active zone enters Australia at the port Derby.
}
\label{scenarioA}
\end{figure}

Two examples shown in Fig.~\ref{GlobesTrajectories}. demonstrate that colonisation of East and South Asia 
can go by different scenarios however propagations towards Western Europe and Americas have matching trajectories.
Let us consider scenario in Fig.~\ref{GlobesTrajectories}a. In the second day after inoculation plasmodium propagates 
from \One to \Two and then to \Three, and from \One to \Ten and to \Eight, and from \Ten to \Eleven and 
to \Nine.  On the third day, plasmodium propagates from \Eleven to \Twelve to \Fourteen to \Thirteen
(Fig.~\ref{6417}). It then continuous to explore space north-east of \Thirteen.
On the fourth day plasmodium develops protoplasmic links from \Ten to \Eight, from \Eight to \Five and \Six, and
from \Six to \Seven; and, completes the route from \Thirteen to \One. It also explores space east of \Seven (Fig.~\ref{6441}).  On the fifth, sixth and seventh days slime mould propagates from \Seven to \Twentythree (Fig.~\ref{6455}), and from \Fourteen to \Fifteen. The plasmodium reaches \Eighteen from \Fifteen via Iceland, Greenland and Canada on ninth day after inoculation in \One. Then it propagates from \Eighteen to \Nineteen, \Twenty, \Twentyone and \Twentytwo (Fig.~\ref{GlobesTrajectories}a).

\begin{figure}[!tbp]
\centering
\subfigure[]{\includegraphics[width=0.49\textwidth]{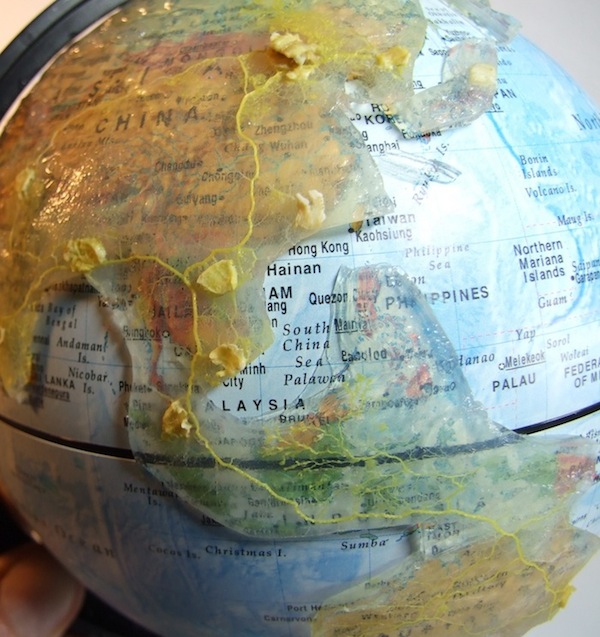}\label{exm6521}}
\subfigure[]{\includegraphics[width=0.49\textwidth]{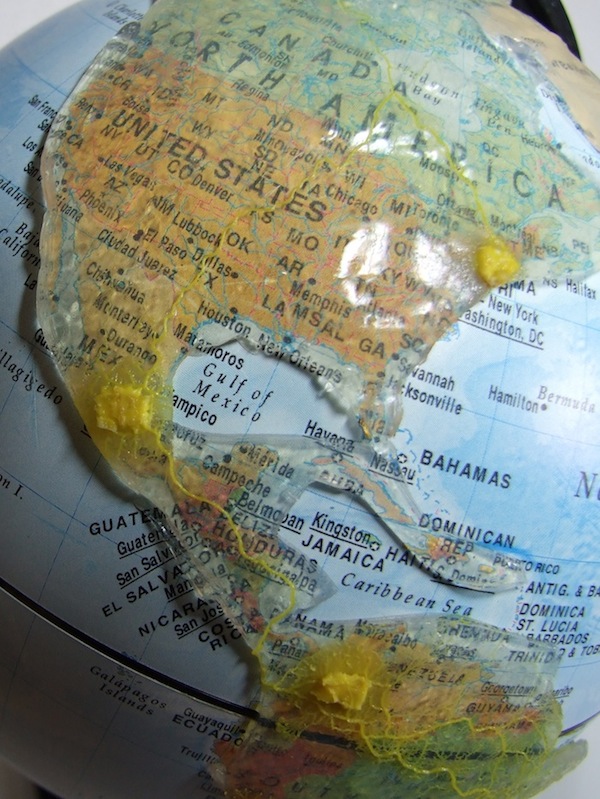}\label{exm6599A}}
\subfigure[]{\includegraphics[width=0.49\textwidth]{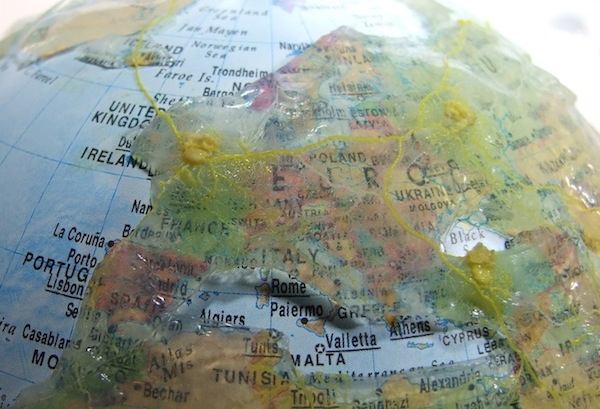}\label{exm6565A}}
\subfigure[]{\includegraphics[width=0.49\textwidth]{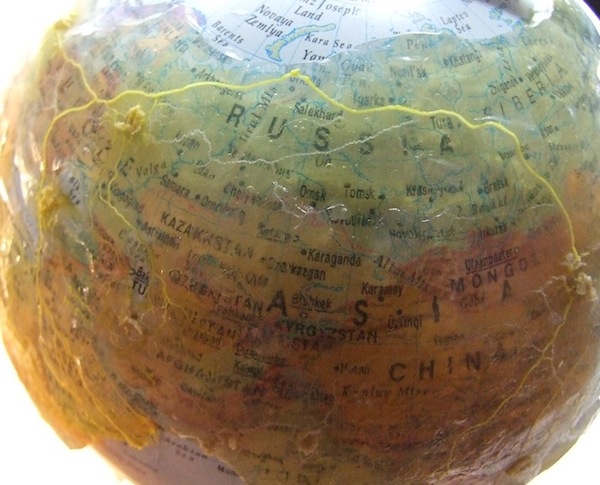}\label{exm6561A}}
\caption{Illustrations of the scenario Fig.~\ref{GlobesTrajectories}b.
\subref{exm6521} Plasmodium propagates from South-East Asia to Australia.
\subref{exm6565A} Plasmodium propagates west and north-west.
\subref{exm6599A} Plasmodium reaches USA via Iceland, Greenland and Canada and then spans Latin America.
\subref{exm6561A} Transport link between \One and \Fifteen.
} 
\label{scenarioB}
\end{figure}

Early stages of colonisation (Fig.~\ref{GlobesTrajectories}a) show the following priorities of the plasmodium's development: east then west then south. The scenario with early development: east then south then west, is shown in
Fig.~\ref{GlobesTrajectories}b and illustrated in Figs.~\ref{exm6521}, \ref{exm6565A}, \ref{exm6599A}, 
\ref{exm6561A}. 

In first two days after inoculation in \One the plasmodium propagates to \Two and \Three in the east, to \Ten in 
the south-west  and \Four in the south-east. It also ventures north to Siberia.  On the third day the plasmodium 
grows from \Five to \Eight and \Six,  from \Six to \Seven  and \Twentythree (Fig.~\ref{exm6521}). On the forth day the 
plasmodium propagates from \Ten to \Eleven, \Twelve, \Fourteen, \Thirteen, \Fifteen, and ventures from \Fifteen to 
Iceland (Fig.~\ref{exm6565A}). Protoplasmic links between \Fifteen and \Sixteen and \Seventeen are developed by the slime mould at the fifth day of the experiment. On the sixth and seventh days plasmodium crosses Greenland and Canada towards USA. It firstly reaches \Eighteen  and then propagates towards \Nineteen, from where it spans \Twenty, 
\Twentyone and \Twentytwo  (Fig.~\ref{exm6599A}). This scenarios also displays a pronounced transport link between 
\One and \Fifteen (Fig.~\ref{exm6561A}).


\begin{figure}[!tbp]
\centering
\includegraphics[width=0.8\textwidth]{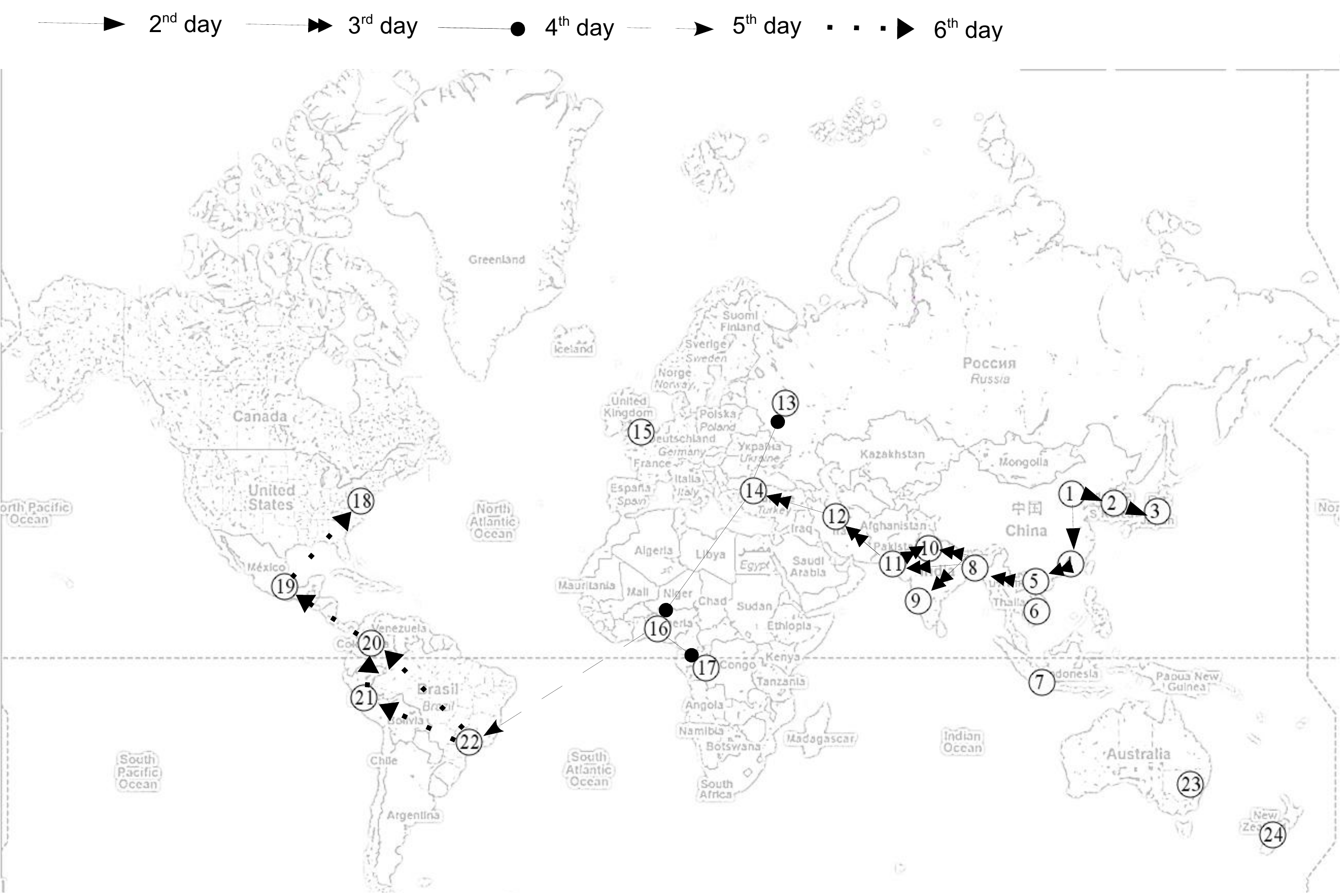}
\caption{Examples of incomplete spanning of urban areas. 
Trajectories of plasmodial active zone propagation in ten days of experiment.}
\label{GlobesTrajectories1}
\end{figure}

\begin{figure}[!tbp]
\centering
\subfigure[]{\includegraphics[width=0.49\textwidth]{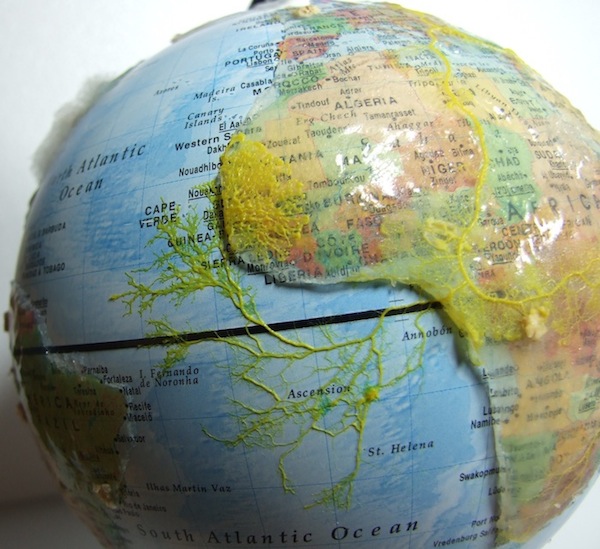}\label{exm6316A}}
\subfigure[]{\includegraphics[width=0.49\textwidth]{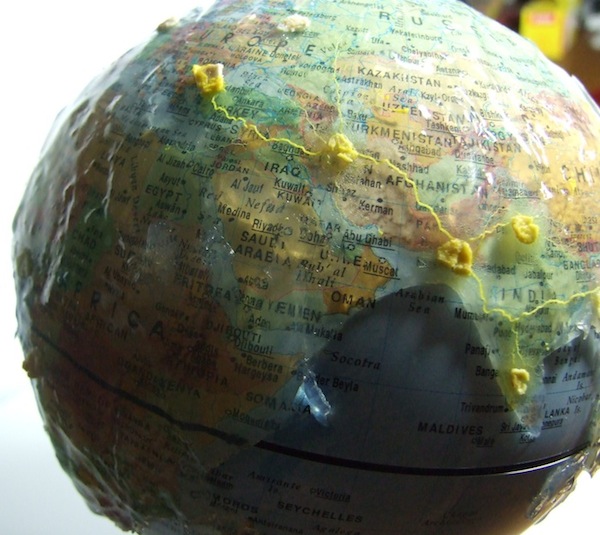}\label{exm6216A}}
\subfigure[]{\includegraphics[width=0.39\textwidth]{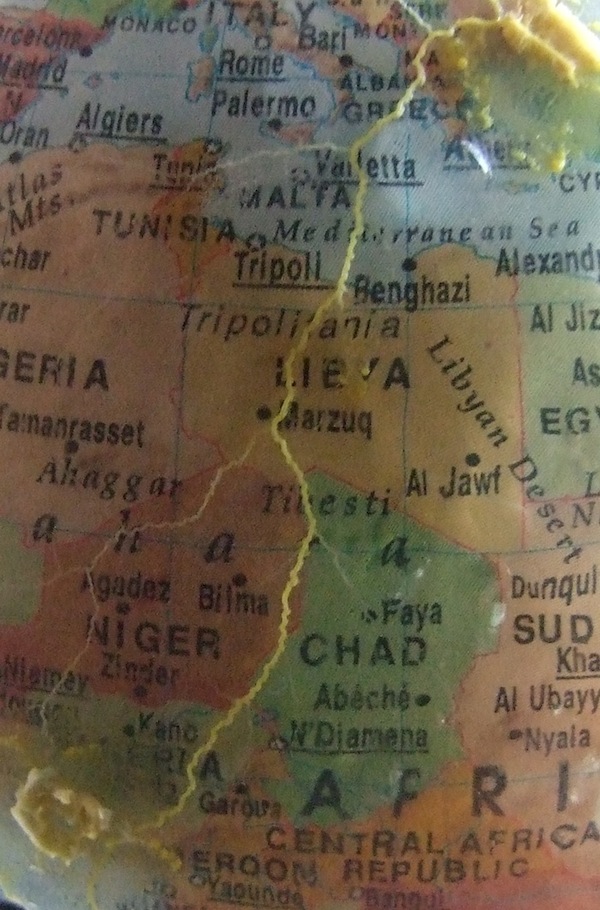}\label{exm6263A}}
\subfigure[]{\includegraphics[width=0.49\textwidth]{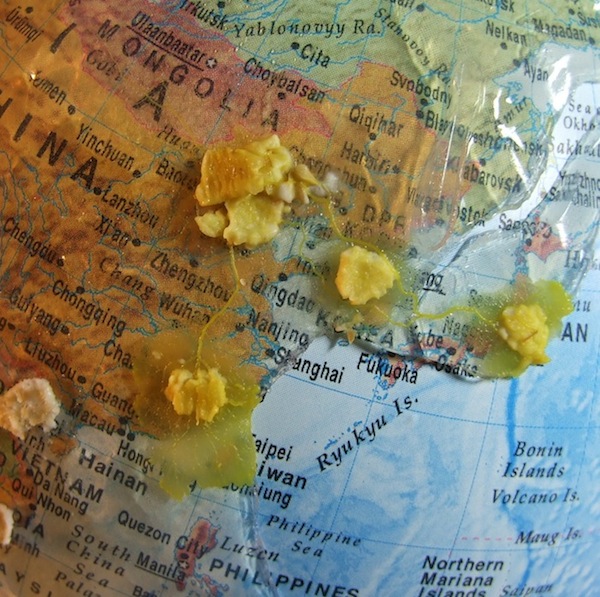}\label{exm6190A}}
\caption{
Illustration of the scenario Fig.~\ref{GlobesTrajectories1}.
\subref{exm6316A}  Plasmodium crosses South Atlantic from West Africa to South America.
\subref{exm6263A} Protoplasmic tube connecting \Fourteen to \Sixteen develops on the fourth day of experiment.
\subref{exm6190A} Transport links \One, \Two and \Three, and \One and \Four  are established in two days after inoculation. 
\subref{exm6216A} Chain \Four -- \Five -- \Eight -- \Ten -- \Eleven -- \Twelve -- \Fourteen is developed on the third day of experiment.
}
\label{scenario2}
\end{figure}





Example of incomplete spanning is shown in Fig.~\ref{GlobesTrajectories1}. Transport links between \One, \Two and \Three, and \One and \Four  are established in two days after inoculation (Fig.~\ref{exm6190A}).
On the third day the plasmodium connects the following areas by the chain of protoplasmic tubes 
\Four -- \Five -- \Eight -- \Ten -- \Eleven -- \Twelve -- \Fourteen  (Fig.~\ref{exm6216A}). \Thirteen and \Sixteen and \Seventeen are spanned by the protoplasmic network on the fourth day of colonisation  (Fig.~\ref{exm6263A}). It takes slime mould one more day to cross South Atlantic to make the link between \Sixteen and \Twentytwo (Fig.~\ref{exm6316A}).  Protoplasmic links (\Twentytwo, \Twentyone), (\Twentytwo, \Twenty),  (\Twenty, \Nineteen), 
(\Nineteen, \Eighteen) are grown on the sixth day of the experiment (Fig.~\ref{GlobesTrajectories1}.).

\begin{figure}[!tbp]
\centering
\subfigure[]{\includegraphics[width=0.49\textwidth]{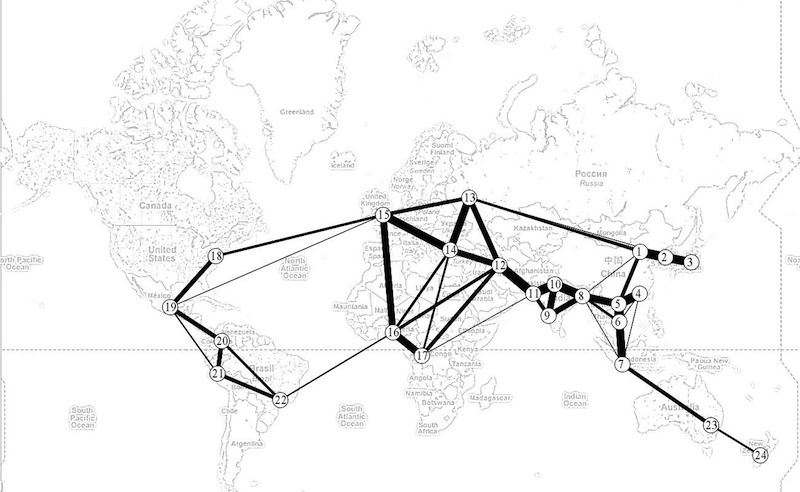}}
\subfigure[]{\includegraphics[width=0.49\textwidth]{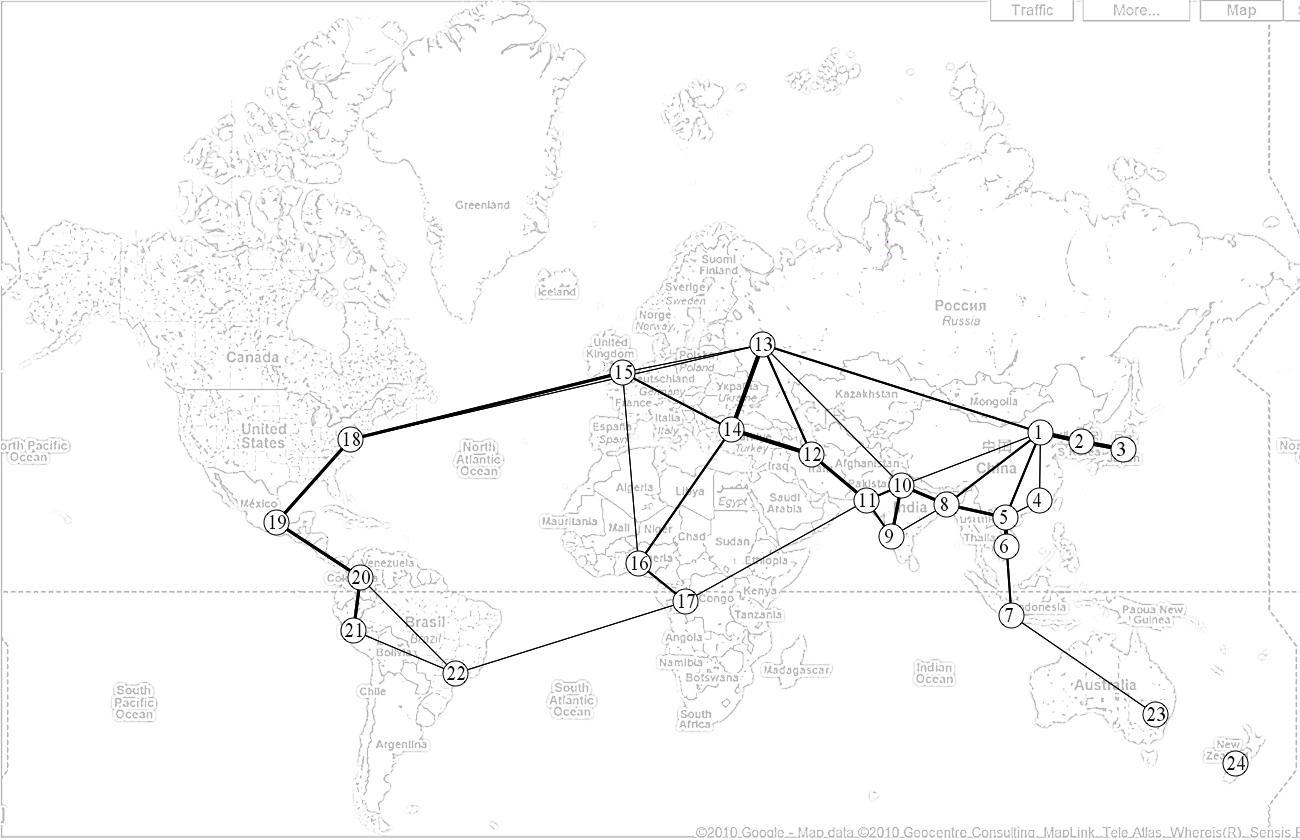}}
\caption{Physarum graphs  obtained from 
experiments in (a)~the Petri dishes, P-graph, and (b)~the globe, G-graph. 
Thickness of a line is proportional to the line's weight.}
\label{rawgraphs}
\end{figure}

Generalised Physarum graphs derived from experiments in Petri dishes (let us call them P-graphs) 
and the globe (G-graphs) are shown in Fig.~\ref{rawgraphs}.  P-graphs represent G-graphs sufficiently.  
Only three edges (albeit represented in one experiment each) of P-graphs are not represented by G-graphs: 
(\Seventeen, \Twentytwo),  (\Thirteen, \Fourteen) and (\One, \Four).  In contrast, eight edges of P-graphs are not
represented by G-graphs: (\Four, \Seven), (\Five, \Six), (\Seven, \Eight), (\Twelve, \Sixteen), (\Fourteen, \Seventeen), (\Fifteen, \Nineteen), (\Nineteen, \Twenty) and (\Sixteen, \Twentytwo). Further in the paper we consider generalised Physarum graphs as a union of P- and G-graphs.
 
 \begin{figure}[!tbp]
\centering
\subfigure[$\theta=\frac{1}{38}$]{\includegraphics[width=0.43\textwidth]{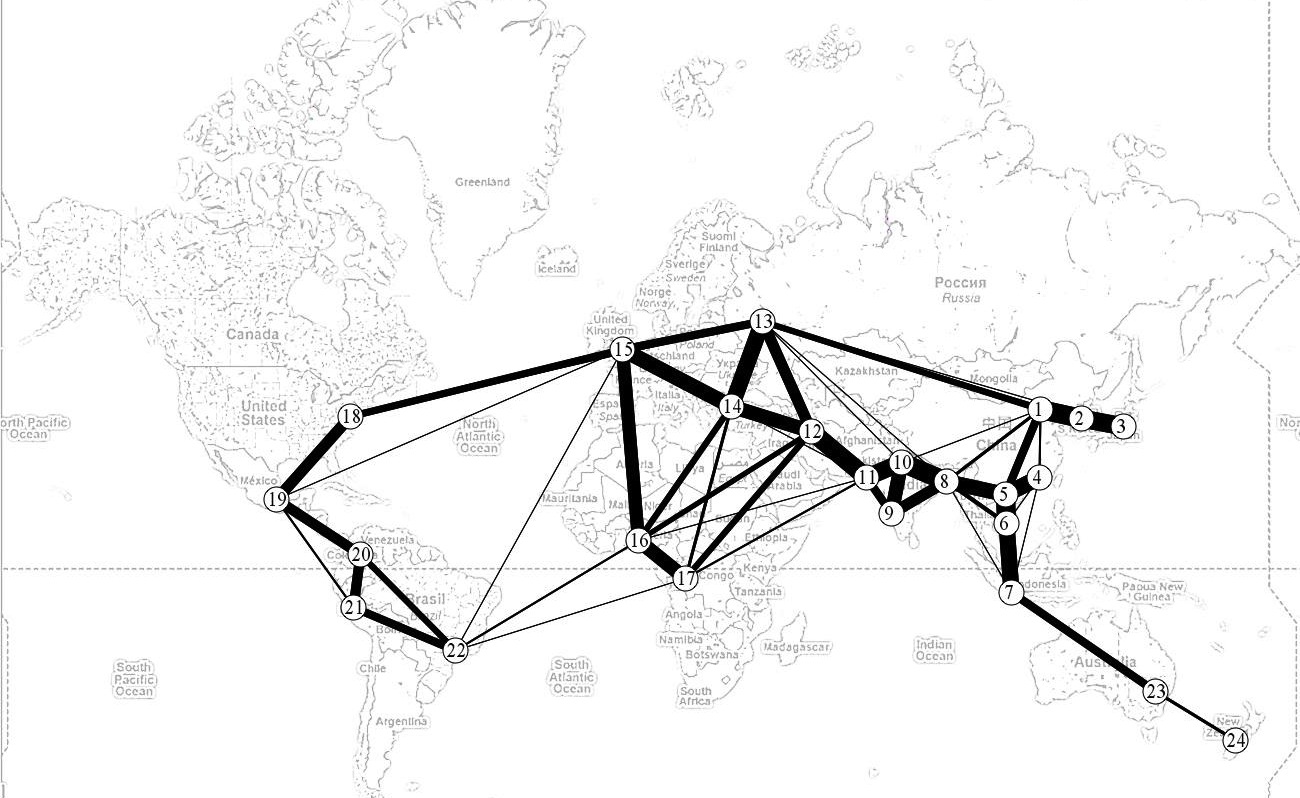}}
\subfigure[$\theta=\frac{6}{38}$]{\includegraphics[width=0.43\textwidth]{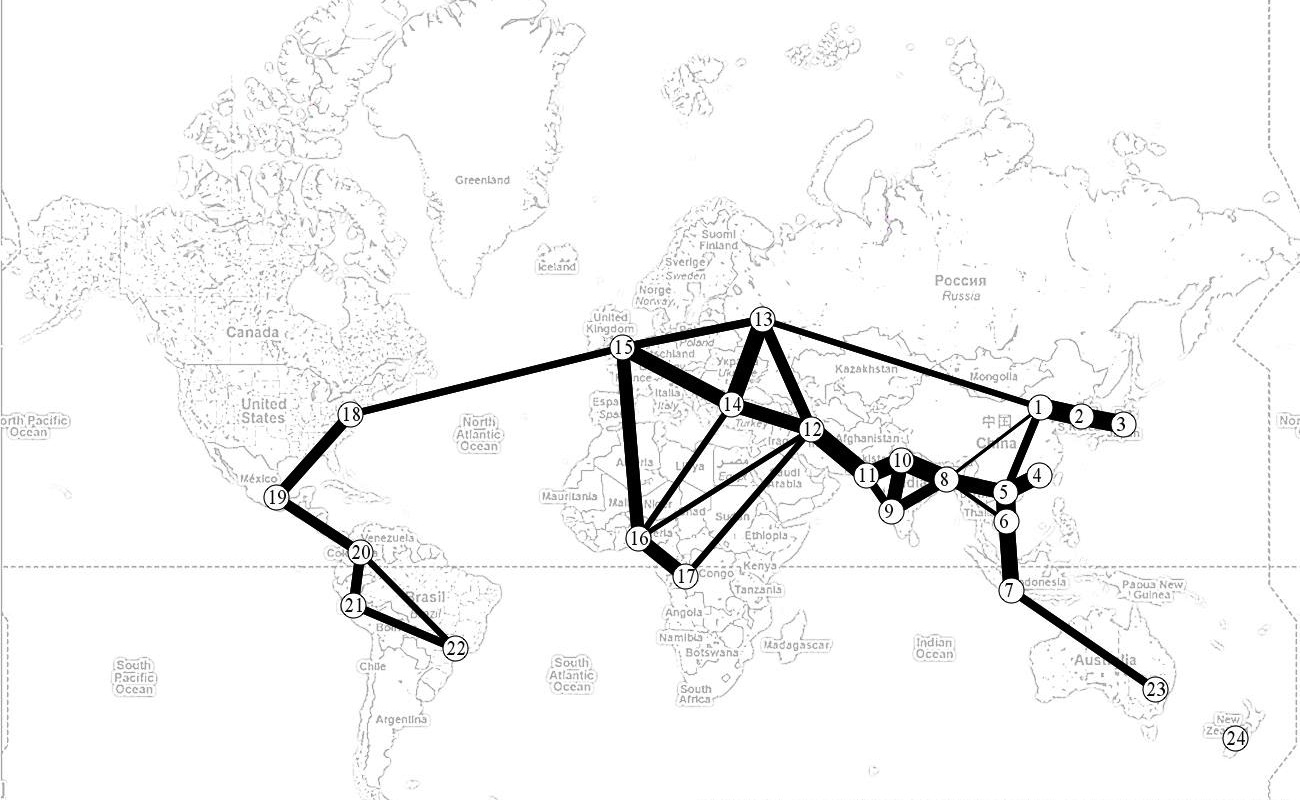}}
\subfigure[$\theta=\frac{14}{38}$]{\includegraphics[width=0.43\textwidth]{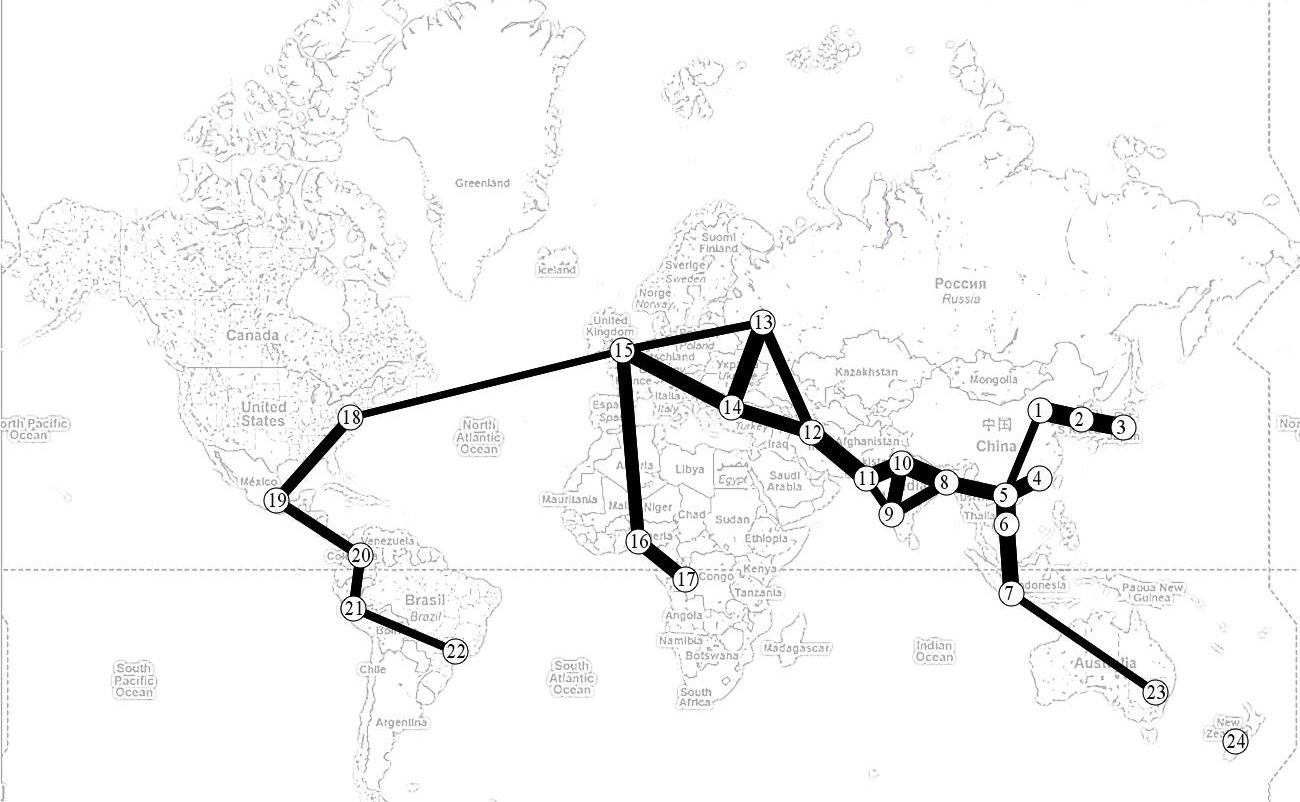}}
\subfigure[$\theta=\frac{15}{38}$]{\includegraphics[width=0.43\textwidth]{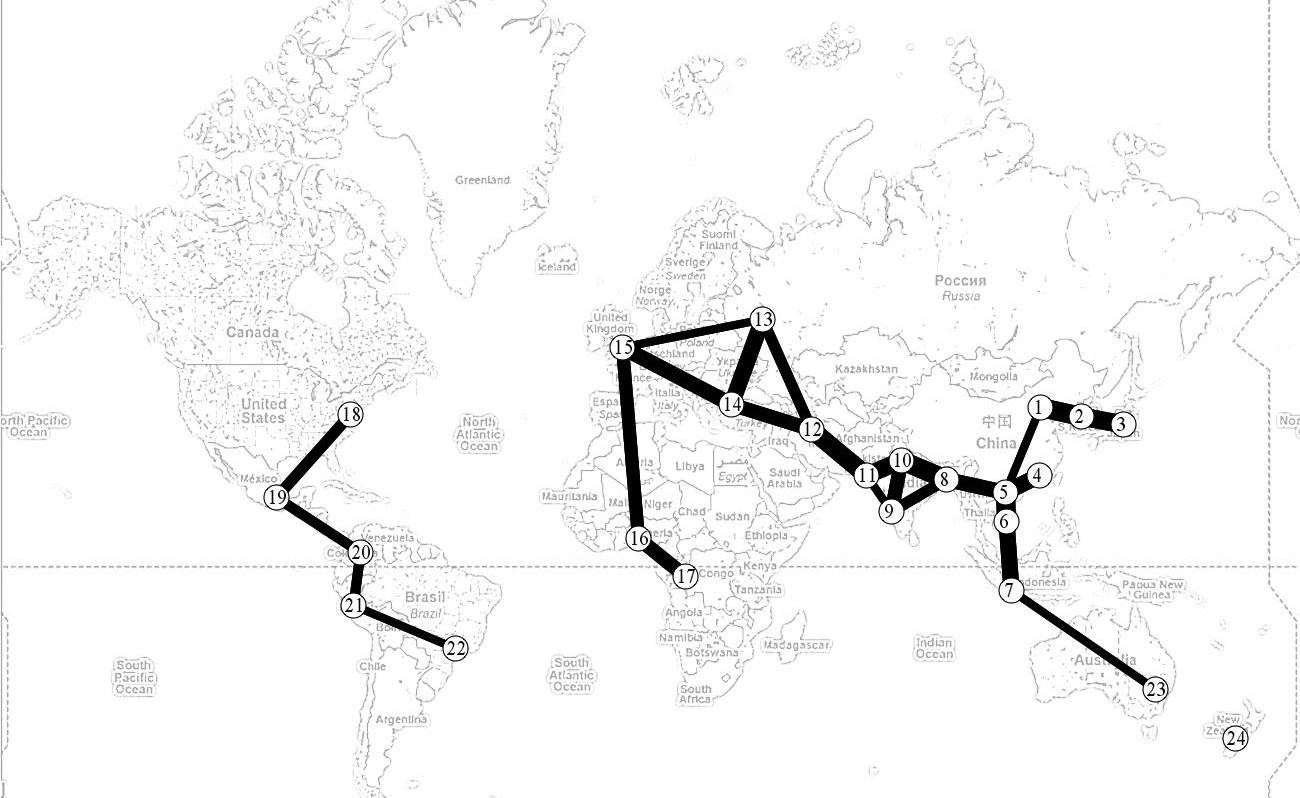}}
\subfigure[$\theta=\frac{16}{38}$]{\includegraphics[width=0.43\textwidth]{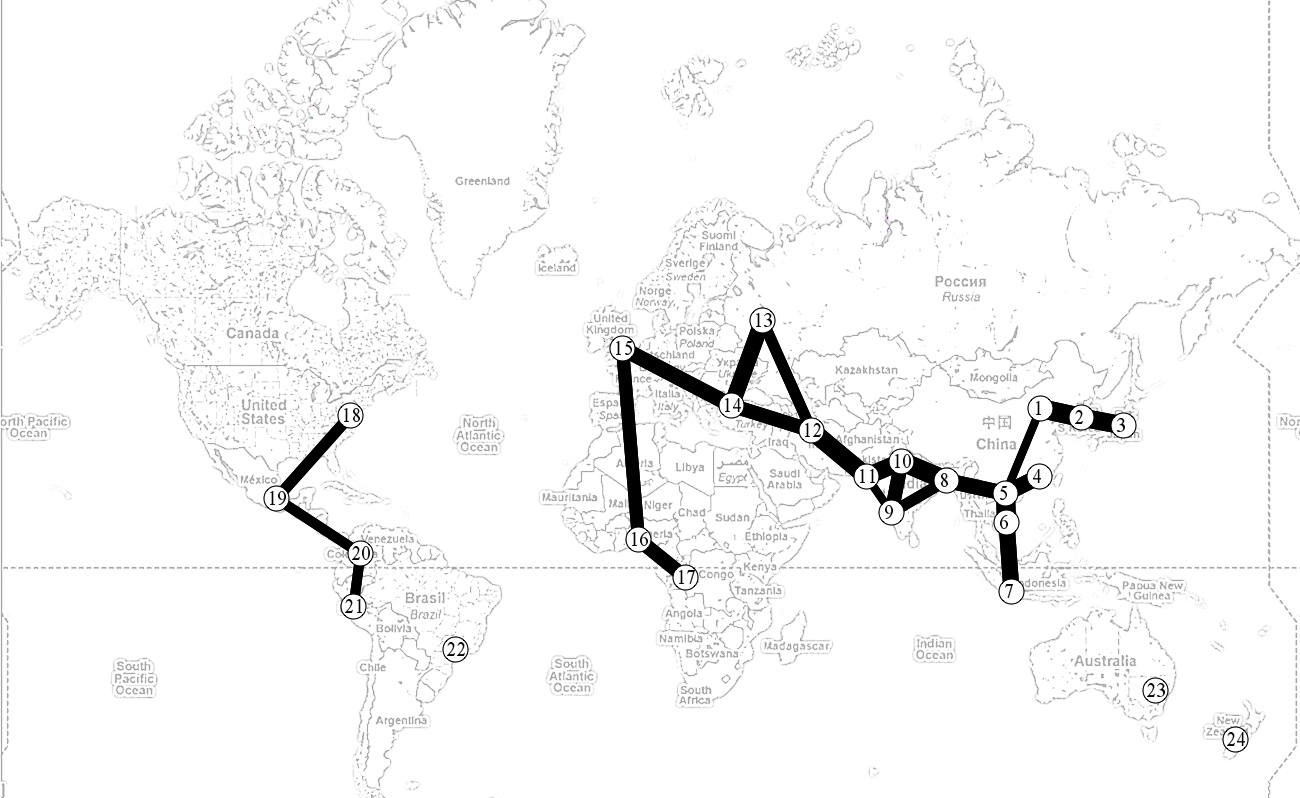}}
\subfigure[$\theta=\frac{22}{38}$]{\includegraphics[width=0.43\textwidth]{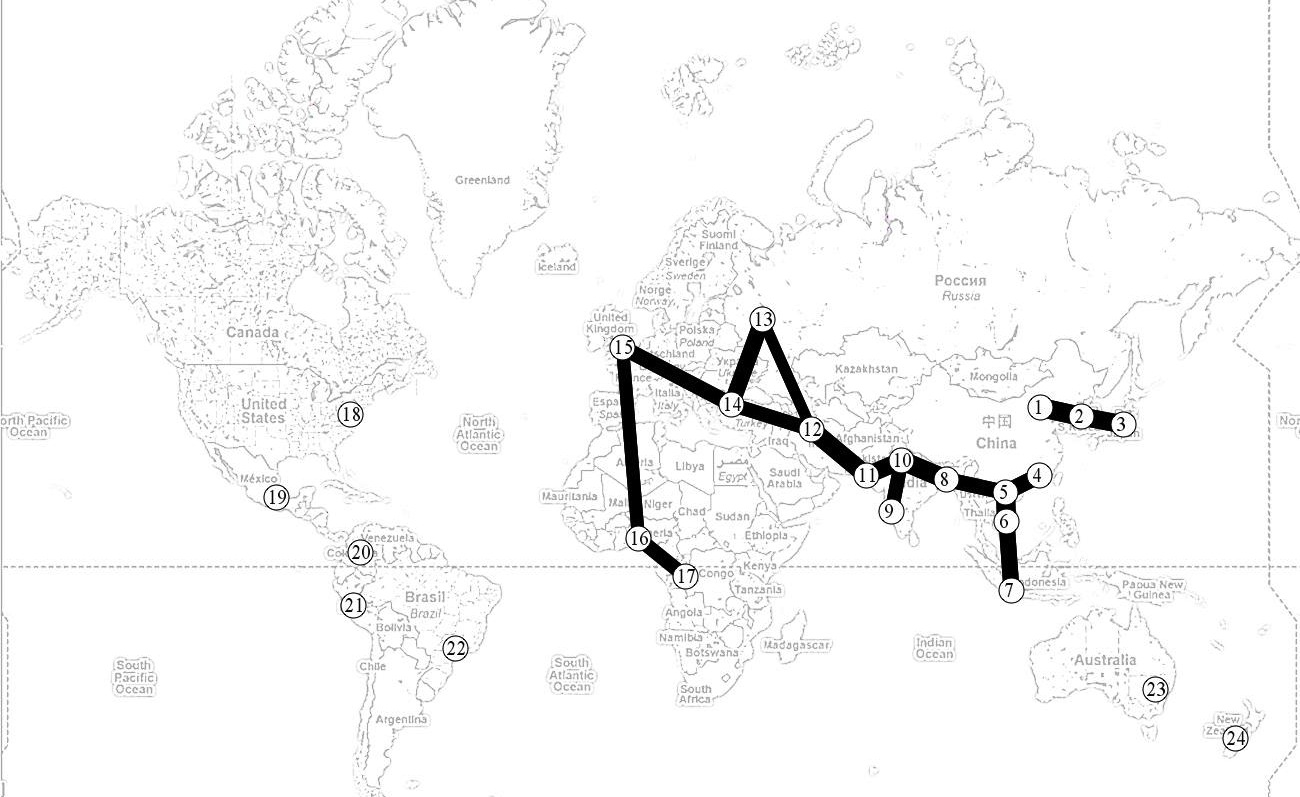}}
\subfigure[$\theta=\frac{23}{38}$]{\includegraphics[width=0.43\textwidth]{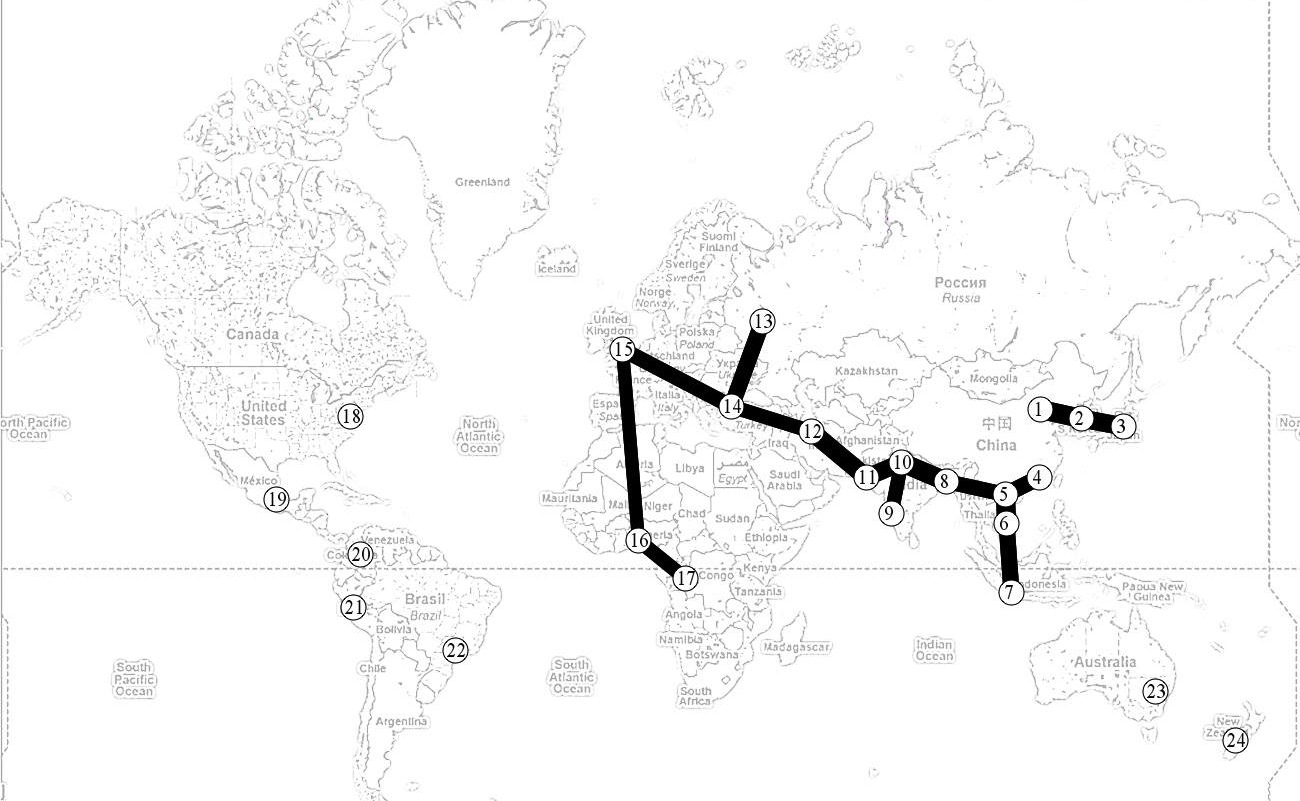}}
\caption{Generalized Physarum graphs $\mathbf{P}(\theta)$ for selected values of $\theta$.}
\label{physarumgraphs}
\end{figure}
 
Generalised Physarum graphs for critical values of $\theta$ are shown in Fig.~\ref{rawgraphs}.  Generalised Physarum graphs $\mathbf{P}(\theta)$ are sub-graphs of P- and G-graphs for $\theta \geq \frac{14}{38}$.
Physarum graphs are planar for $\theta \geq \frac{6}{38}$ however with acquiring planarity the Physarum graph 
$\mathbf{P}(\frac{6}{38})$ loses connectivity and \Twentyfour becomes isolated (Fig.~\ref{rawgraphs}b).  The highest value of $\theta$ for which graph $\mathbf{P}(\theta)-\{$\Twentyfour$\}$ remains connected is $\frac{14}{38}$ 
 (Fig.~\ref{rawgraphs}c). When $\theta$ increases to $\frac{15}{38}$ Americas become disconnected from Eurasia 
 and Africa, and American transport pathways become a single chain \Eighteen --  \Nineteen -- \Twenty -- \Twentyone -- \Twentytwo  (Fig.~\ref{rawgraphs}d).  Further increase of $\theta$ to $\frac{16}{38}$ leads to isolation of \Twentytwo and \Twentythree (Fig.~\ref{rawgraphs}e).
 
 For $\theta=\frac{22}{38}$ all Americas areas of $\mathbf{U}$ become isolated and the transport network at this continent virtually disappears. The graph  $\mathbf{P}(\frac{22}{38})$ has two components of connectivity: 
 \begin{itemize}
 \item the chain \One -- \Two -- \Three
 \item the component consisting of the chain  \Seven  -- \Six -- \Eight -- \Ten -- \Twelve -- \Fourteen -- \Fifteen -- \Sixteen -- \Seventeen with two branches \Five -- \Four and \Ten -- \Nine attached and the cycle \Twelve -- \Thirteen -- \Fourteen -- \Twelve  (Fig.~\ref{rawgraphs}f).
 \end{itemize}
 
 Generalised Physarum graph is transformed to the acyclic graph at $\theta=\frac{23}{38}$ when the edge (\Twelve, \Thirteen) becomes cut off (Fig.~\ref{rawgraphs}g.)

\subsection{Proximity graphs}

\begin{figure}[!tbp]
\centering
\subfigure[$\mathbf{MST}$]{\includegraphics[width=0.49\textwidth]{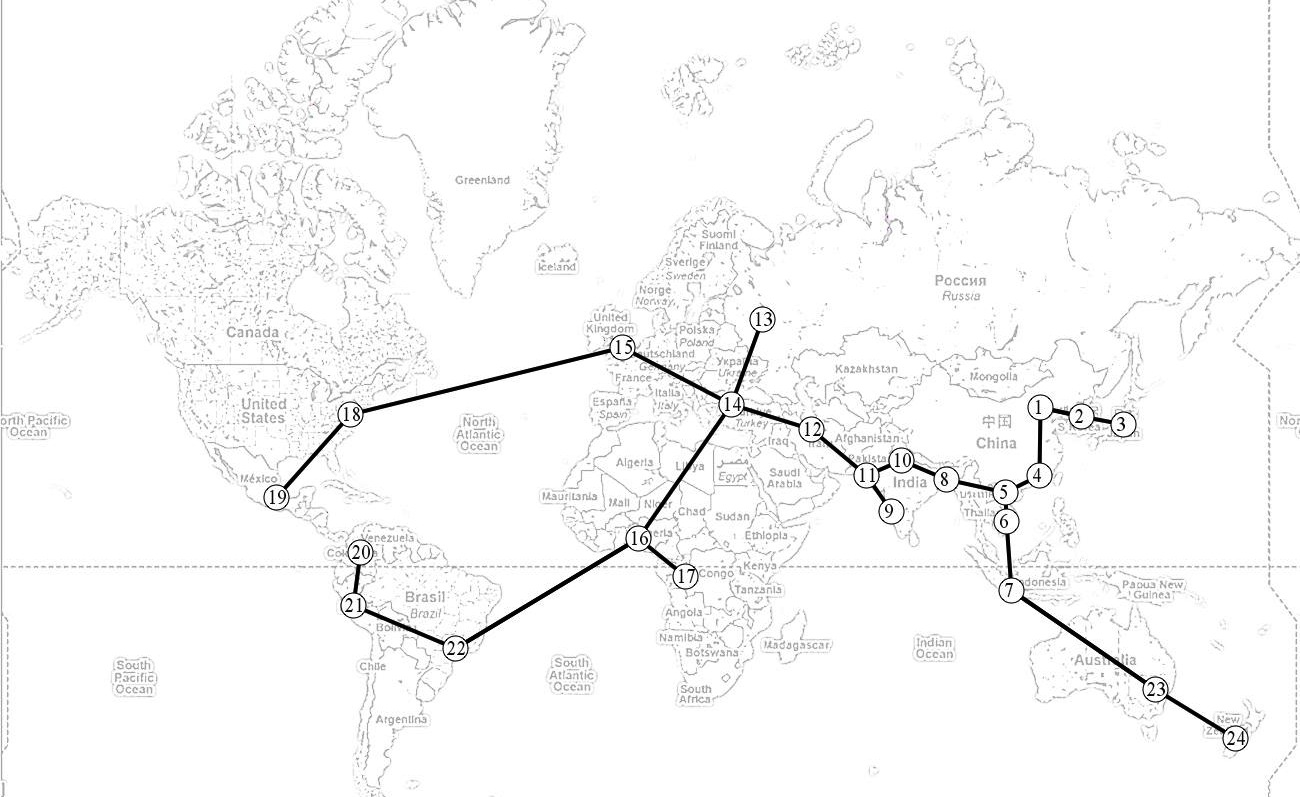}}
\subfigure[$\mathbf{RNG}$]{\includegraphics[width=0.49\textwidth]{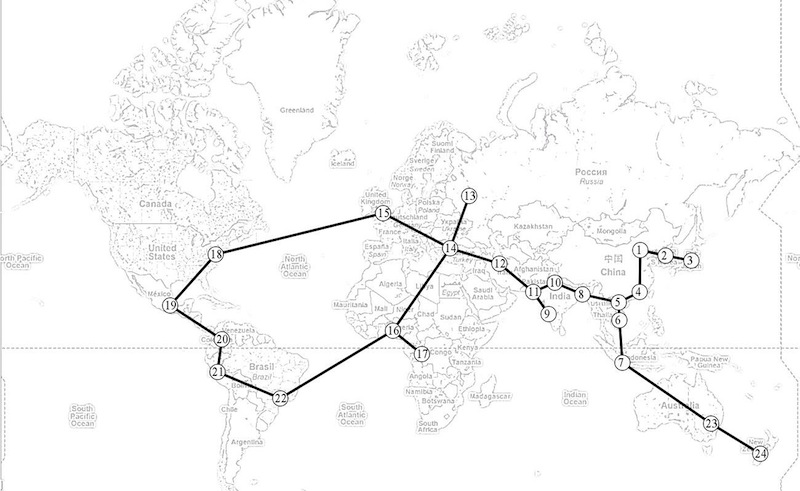}}
\subfigure[$\mathbf{GG}$]{\includegraphics[width=0.49\textwidth]{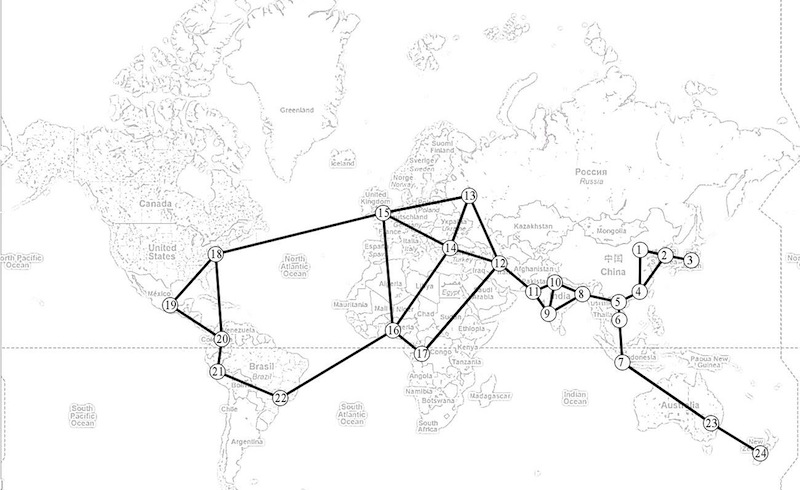}}
\caption{Relative neighbourhood graph~(a) and Gabriel graph~(b) constructed on urban areas of $\mathbf{U}$.}
\label{proximity}
\end{figure}

\begin{figure}[!tbp]
\centering
\subfigure[$\mathbf{MST} \cap \mathbf{P}(\frac{1}{38})$]{\includegraphics[width=0.49\textwidth]{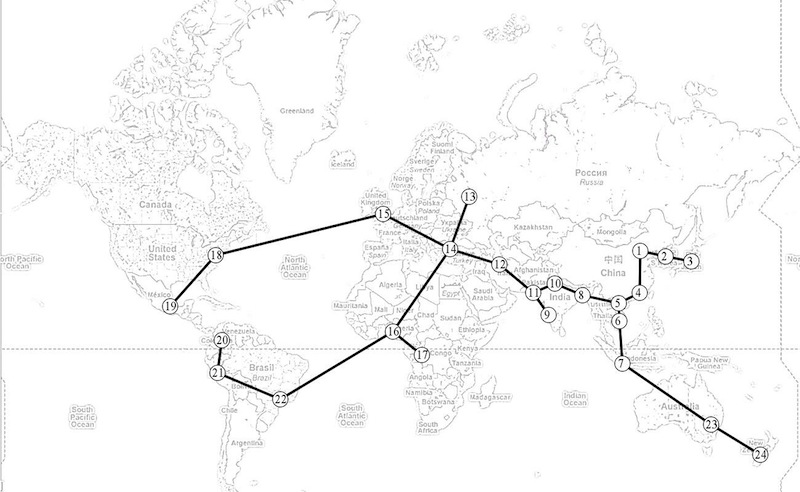}}
\subfigure[$\mathbf{RNG} \cap \mathbf{P}(\frac{1}{38})$]{\includegraphics[width=0.49\textwidth]{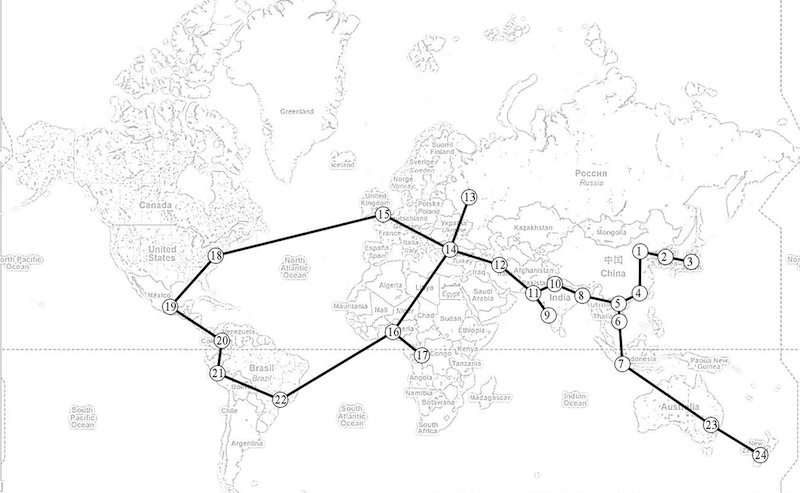}}
\subfigure[$\mathbf{GG} \cap \mathbf{P}(\frac{1}{38})$]{\includegraphics[width=0.49\textwidth]{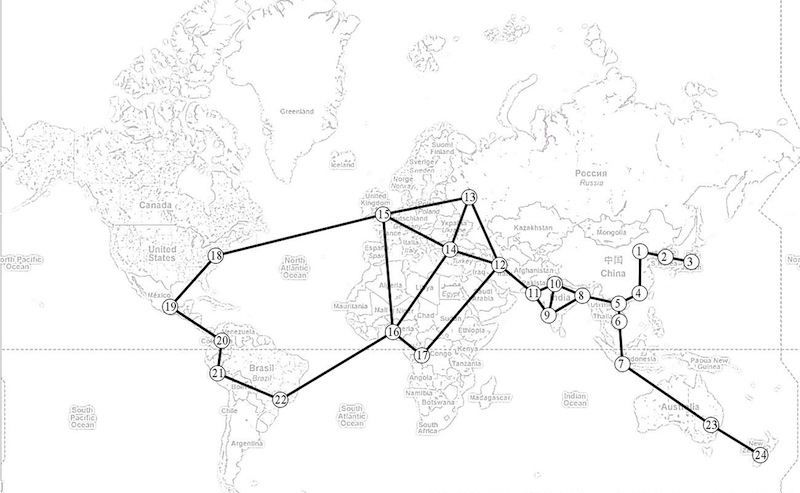}}
\subfigure[$\mathbf{MST} \cap \mathbf{P}(\frac{14}{38})$]{\includegraphics[width=0.49\textwidth]{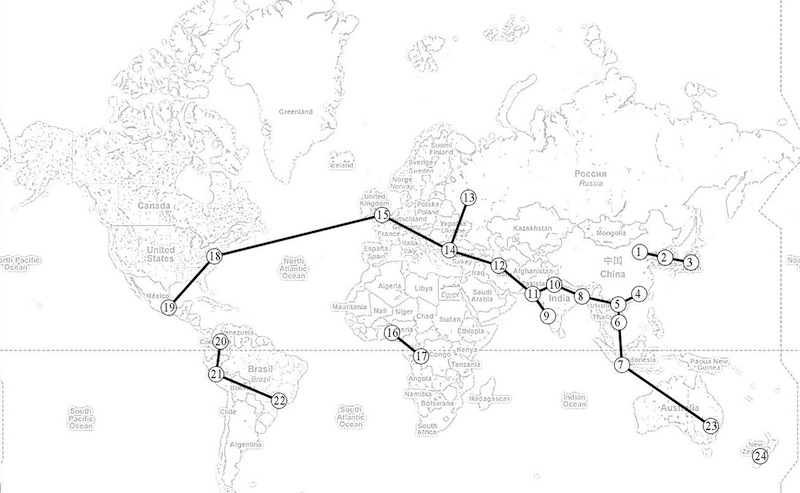}}
\subfigure[$\mathbf{RNG} \cap \mathbf{P}(\frac{14}{38})$]{\includegraphics[width=0.49\textwidth]{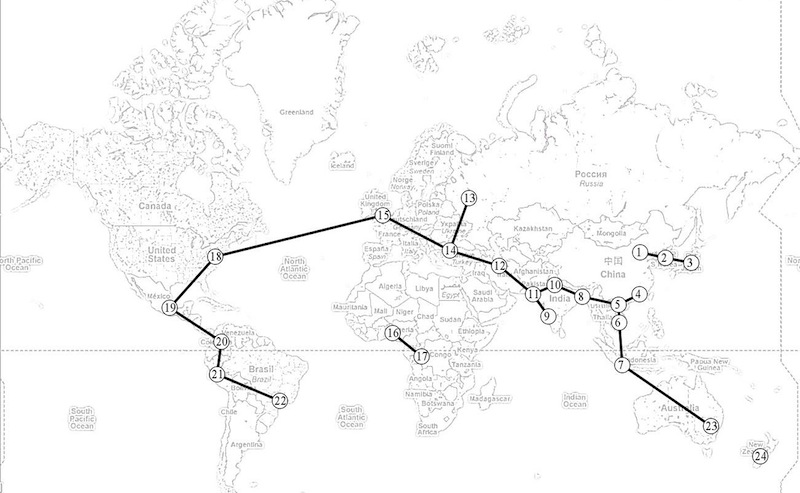}}
\subfigure[$\mathbf{GG} \cap \mathbf{P}(\frac{14}{38})$]{\includegraphics[width=0.49\textwidth]{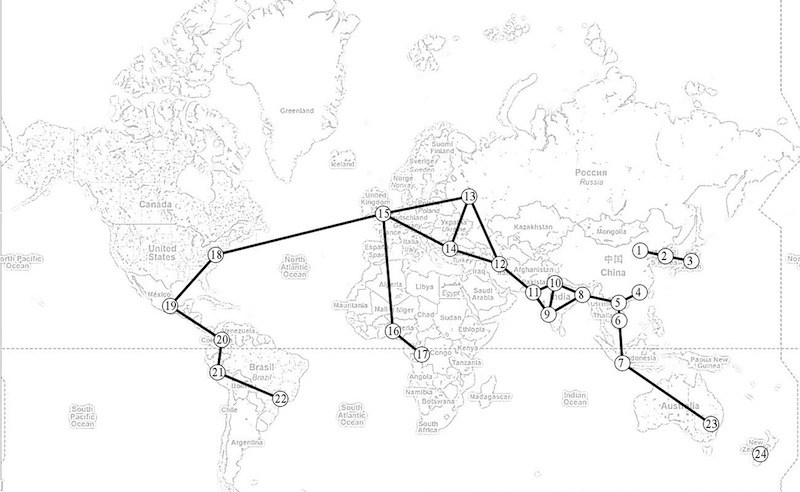}}
\caption{Intersections of generalised Physarum graphs (abc)~$\mathbf{P}(\frac{1}{38})$ and 
(def)~$\mathbf{P}(\frac{14}{38})$ with proximity graphs: 
(ad)~minimum spanning tree $\mathbf{MST}$, 
(be)~relative neighbourhood graph $\mathbf{RNG}$, 
(cf)~Gabriel graph $\mathbf{GG}$.}
\label{RNGGGPhys}
\end{figure}

A planar graph consists of nodes which are points of the  Euclidean plane and edges which are straight segments connecting the points. A planar proximity graph is a planar graph where two points are connected by an edge if they are close in some sense. A pair of points is assigned a certain neighbourhood, and points of the pair are connected by an edge if their neighbourhood is empty.  Here we consider the most common proximity graphs as follows.

$\mathbf{MST}$: The Euclidean minimum spanning tree~\cite{nesetril} is a connected acyclic graph which has minimum possible sum of edges' lengths (Fig.~\ref{proximity}a).

$\mathbf{RNG}$: Points $a$ and $b$ are connected by an edge in the Relative Neighbourhood Graph if no other point $c$ is closer to $a$ and $b$ than $dist(a,b)$~\cite{toussaint_1980} (Fig.~\ref{proximity}b).

$\mathbf{GG}$: Points $a$ and $b$ are connected by an edge in the Gabriel Graph if
disc with diameter $dist(a,b)$ centred in middle of the segment $ab$ is empty~\cite{gabriel_sokal_1969, matula_sokal_1984} (Fig.~\ref{proximity}a).

In general, the graphs relate as $\mathbf{MST} \subseteq \mathbf{RNG}  \subseteq\mathbf{GG}$~\cite{toussaint_1980,matula_sokal_1984,jaromczyk_toussaint_1992}; this is called Toussaint hierarchy.

Why do we need to compare Physarum graphs with the proximity graphs $\mathbf{MST}$, $\mathbf{RNG}$ and  
$\mathbf{GG}$? The minimum spanning tree helps to evaluate optimality of the protoplasmic networks: minimal distances of nutrient transportation yet complete covering of 
the sources of nutrients (sites of $\mathbf{U}$).  Being an acyclic graph the spanning tree is sensitive to a structural damage. Removal of a single edge might transform the spanning tree to two disconnected trees. The relative neighbourhood graph and the Gabriel graph show higher degree of fault-tolerance (depending on exact configuration of nodes). It also considered to be optimal in terms of total edge length and travel distance.  The graphs $\mathbf{RNG}$ and $\mathbf{GG}$ are used in geographical variational analysis~\cite{gabriel_sokal_1969,matula_sokal_1984},  simulation of epidemics~\cite{toroczkai_2008}, and design of \emph{ad hoc} wireless  networks~\cite{li_2004,song_2004,santi_2005,muhammad_2007,wan_2007}. The proximity graphs, especially 
$\mathbf{RNG}$, are invaluable in simulation of human-made, road networks; these graphs are validated in studies of Tsukuba central district road networks~\cite{watanabe_2005, watanabe_2008}. Gabriel graphs, particularly their relaxed versions~\cite{bose_2009} are an ideal tool for path finding and online routing on planar graphs~\cite{bose_2004}.

Intersections of generalised Physarum graphs with the proximity graphs is shown in (Fig.~\ref{RNGGGPhys}). 

\begin{finding} 
The  minimum spanning tree and the relative neighbourhood graph are sub-graphs of
 $\mathbf{P}(\frac{1}{38})$. Gabriel graph would be a sub-graph of  $\mathbf{P}(\frac{1}{38})$ if the Physarum graph had the edges (\Four, \Two) and (\Eighteen, \Twenty).
\end{finding}

The $\mathbf{MST}$ differs from $\mathbf{RNG}$ only by absence of the single edge (\Nineteen, \Twenty), the rest is the same and is matched by edges of the generalised Physarum graph $\mathbf{P}(\frac{1}{38})$ 
(Fig.~\ref{RNGGGPhys}ab).  The fact that $\mathbf{GG} \subseteq \mathbf{P}(\frac{14}{38}) \cup$(\Four, \Two)$\cup$(\Nineteen, \Twenty) (Fig.~\ref{RNGGGPhys}c) plays against the geographical validity of the  Gabriel graph. The only physically possible routes from \Eighteen to \Twenty and from \Four to \Two are maritime routes; overland route from 
\Eighteen to \Twenty is passing via \Nineteen and from \Four to \Two via \One.  

\begin{finding}
The longest chain in $\mathbf{MST} \cap \mathbf{P}(\frac{14}{38})$ is 
$C_1$=(\Twentythree -- \Seven -- \Six -- \Five -- \Eight -- \Ten -- \Eleven -- \Twelve -- \Fourteen -- \Fifteen -- \Eighteen -- \Nineteen) 
\end{finding}

Two fragments of the chain $C_1$: \Seven -- \Six -- \Five and \Eight to \Fifteen (Fig.~\ref{RNGGGPhys}d)
match the Silk Road, see Sect.~\ref{silk}. The chain $C_1$  is elongated by segment (\Twenty -- \Twentyone -- \Twentytwo) in $\mathbf{RNG} \cap \mathbf{P}(\frac{14}{38})$ (Fig.~\ref{RNGGGPhys}e)  and 
$\mathbf{GG} \cap \mathbf{P}(\frac{14}{38})$ (Fig.~\ref{RNGGGPhys}f).

\subsection{Protoplasmic Networks, the Silk Road and the Asian Highways}
\label{silk}

\begin{figure}[!tbp]
\centering
\subfigure[]{\includegraphics[width=0.49\textwidth]{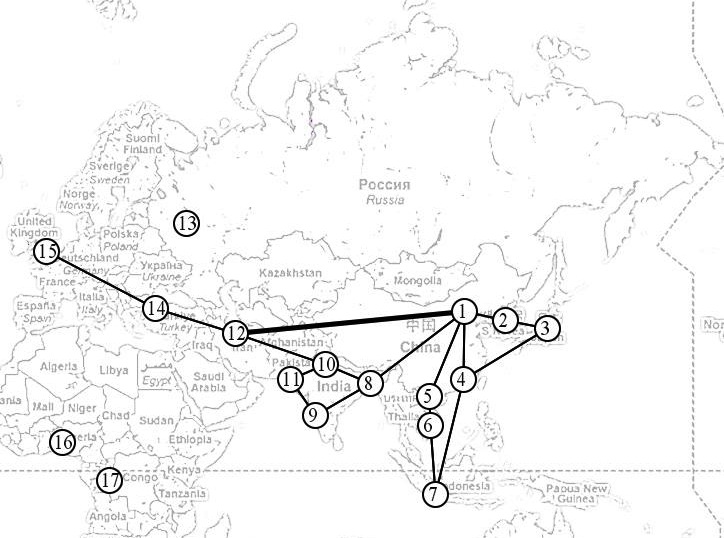}}
\subfigure[]{\includegraphics[width=0.49\textwidth]{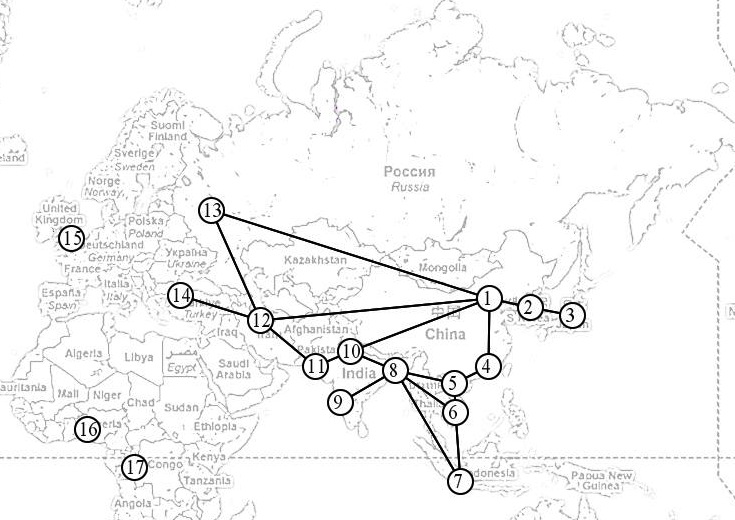}}
\caption{Graphs of the Silk Road and the Asian Highways. 
(a)~the Silk Road graph, comprised of the Silk Road (thick line) and associated trade 
routes (thin lines); the graph is derived 
from the scheme of Trans-Asian trade 500 BC -- AD 750~\cite{ScarreTimesAtlas}.
(b)~the Asian Highway network graph; the graph is derived from the Asian Highway Network 
Map~\cite{Prioritiies}.}
\label{silkandhighways}
\end{figure}

To evaluate how well plasmodial networks match large-scale transportation systems we compare 
generalised Physarum graphs with graphs derived from the Silk Road and the Asia Highway network.

The Silk Road is a long-distance trade route across Asia developed in the first millennium to transport goods 
between China and Western Asia and Northern Europe~\cite{Thubron_2008}.  Commonly the Silk Road is comprised 
of the main overland route from Luoyang to Seleucia/Clesiphon and associated overland and maritime 
routes reaching as far Japan in the east and as far as Colchester and Holborough in the 
west~\cite{ScarreTimesAtlas, Liu_2010}.

A graph of the Silk Road  is shown in Fig.~\ref{silkandhighways}a. 
When constructing the Silk road graph we tracked geographical positions of modern cities along the 
historical Silk Road and associated trade routes, with minor assumptions. Thus the Silk Road is represented by 
edge (\One, \Twelve). The road actually started at Luoyang and passed via Hecatompylos and 
ended in Ctesiphon.  We assumed Hecatompylos was in proximity of \Twelve and \One is in 
proximity of Luoyang. Similarly, we assume that edge \Ten to \Eight corresponds to the trade route connecting 
Mathura (proximity of \Ten) and Tampluk via Pataliputra (proximity of \Eight). Link (\Three, \Four) corresponds to 
a maritime route from south of Japan to Oc Eo via Guangzhou and passing between the continental China and 
Hong Kong island.  The edge (\Twelve,  \Fourteen) represents land route from Hecatompylos to Dura Europos 
and Antioch combined with maritime trade route from Antioch to Constantinple. The edge (\Fourteen, \Fifteen) 
represents maritime route Constantinople -- Athens --  Rome -- Massilia -- Colchester. 

The Asian Highways is a network of roads selected for regional transport cooperation 
``initiative aimed at enhancing efficiency and development of the road transport infrastructure in 
Asia"~\cite{Prioritiies}.  The Asian Highway network consists of 141,000 km of roads running across 
32 member States. When constructing Asian Highway graph (Fig.~\ref{silkandhighways}b) we adopted the following match between the graph's edges and the highways (see map of the Asian Highway network in~\cite{Prioritiies}):
\begin{itemize}
\item (\One, \Two): potential route AH1 Beijing to Shenyang and existing route AH1 Shenyang to Seoul; 
\item (\One, \Four): route AH1  Beijing to Hong Kong via Zhengzhou, Xinyang, Xianglan; 
\item (\One, \Ten):  route A1 Beijing to Zhengzhou, potential routes AH34 Zhengzhou to Xi'an, AH5 Xi'an to Langzhou, 
AH42 Langzhou to Lhasa, and final existing routes AH42 Lhasa to Zhangmu and AH2 Zhangmu to Delhi;
\item (\One, \Twelve): comprised of segments of the following highways (ordered from \One to \Twelve) AH1, AH34, AH5, AH4, AH65, AH62, AH76, AH1;
\item  (\One, \Thirteen): AH3 Beijing to Ulan Ude, AH5 from Ulan Ude to Moscow via Irkutsk, Novosibirsk, Omsk, Petropavlosk, Chelyabinsk, Samar;
\item (\Two, \Three): maritime route Pusan to Fukuoka and overland route   AH1 Fukuoka to Tokyo;
\item (\Four, \Five): potential route AH1 Guangzhou to Nanning, AH1 Nanning to Hanoi;
\item (\Five, \Six): AH1; 
\item(\Five, \Eight): AH14 Hanoi -- Kummig -- Mandalay and AH1 Mandalay -- Dispur -- Kolkata;
\item (\Six, \Seven): AH1 \Six to Kabin Bun, AH19 Kabin Bun to Bangkok, AH2 Bangkok to Hat Yai, AH18 Hat Yai to Singapore, and maritime Singapore to Jakarta;
\item (\Eight, \Nine): AH46 and AH47; 
\item (\Eight, \Ten): AH1;
\item (\Ten, \Eleven): AH1 Delhi to Lahore, AH2 Lahore to Rohn, AH4 Rohn to Karachi;
\item (\Eleven, \Twelve): AH7 Karachi to Kandahar and AH1 Kandahar -- Dilaram --  Herat -- Zhabzevar -- Tehran;
\item (\Twelve, \Thirteen): AH8 Tehran -- Baku -- Astahan -- Volgograd -- Moscow; 
\item (\Twelve, \Fourteen): AH1 Tehran -- Yerevan -- Ankara and AH5 Ankara to Istanbul.
\end{itemize}

The Asian Highway graph has the same number of edges as the Silk Road graph however they are not isomorphic 
(Fig.~\ref{silkandhighways}).

\begin{finding}
Slime mould P. polycephalum approximates over 76\% of the Silk Road routes and the Asian Highway routes.
\end{finding}

Physarum graph $\mathbf{P}(\frac{6}{38})$ (Fig.~\ref{physarumgraphs}b)  
approximates 13 of 17 edges of the Silk Road graph(Fig.~\ref{silkandhighways}a) and also 13 of 17 edges of the Asian Highway graph (Fig.~\ref{silkandhighways}b). Physarum graph $\mathbf{P}(\frac{14}{38})$  
(Fig.~\ref{physarumgraphs}c)  approximates 11 of 17 edges of the Silk Road graph and also 11 of 17 edges of the Asian Highway graph. 

\begin{finding}
Transport links (\One, \Twelve) and (\One, \Four) of the Silk Road and the Asian Highway network
are never approximated by P. polycephalum.
\end{finding}

The following routes of the Silk Road are never approximated by any Physarum graph: 
(\Ten, \Twelve), (\One, \Four),  (\Three, \Four), (\Four, \Seven).  
Routes (\One, \Twelve) and (\One, \Eight) are approximated by $\mathbf{P}(\frac{6}{38})$
but not $\mathbf{P}(\frac{14}{38})$. 

The following routes of the Asian highways are not approximated by any Physarum graph: 
(\One, \Twelve), (\One, \Four), (\One, \Ten) and (\Seven,  \Eight). 
Routes (\One, \Thirteen) and (\Six, \Eight) are approximated by $\mathbf{P}(\frac{6}{38})$
but not $\mathbf{P}(\frac{14}{38})$.

\section{Conclusions}

To imitate a hypothetical scenario of the World's colonisation and emergence of principal transport networks 
we represented the continents with geometrically-shaped plates of non-nutrient agar and the major metropolitan areas with sources of nutrients and inoculated plasmodium of \emph{P. polycephalum}  in Beijing. We analysed scenarios of the plasmodium propagation and colonisation of the metropolitan areas. We derived weighted generalised Physarum graphs from the protoplasmic networks recorded in the laboratory experiments.  

We found that the Physarum graphs include basic proximity graphs --- the minimum spanning tree, the  relative neighbourhood graph and the Gabriel graph constructed on the metropolitan areas --- as their sub-graphs. The longest chain of transport links presented in the minimum spanning tree and in the Physarum graph with edge weights exceeding 0.36 is \Twentythree -- \Seven -- \Six -- \Five -- \Eight -- \Ten -- \Eleven -- \Twelve -- \Fourteen -- \Fifteen -- \Eighteen -- \Nineteen.  We found that slime mould \emph{P. polycephalum} approximates over 76\% of the Silk Road 
routes and the Asian Highway network. Transport links (\One, \Twelve) and (\One, \Four) of the Silk Road and the Asian Highway network are never approximated by \emph{P. polycephalum}.  

We believe our experimental results will inspire further thoughts, paradigms and approaches for re-evaluation of historical findings on the emergence of ancient roads and will help to design future trans-continental pathways.  The results could be also applied in analysis of pandemics' dynamics~\cite{beck_2008, HaourKnipe_1996} and in shaping unorthodox approaches to prediction of international military conflicts and battlefield operations~\cite{ilachinski}.

\end{document}